\documentclass[12pt]{article}
\usepackage{amsmath}
\usepackage{graphicx,psfrag,epsf}
\usepackage{enumerate}
\usepackage{natbib}
\usepackage{textcomp}
\usepackage[hyphens]{url} 
\usepackage{hyperref}
\usepackage{lmodern}
\providecommand{\tightlist}{%
\setlength{\itemsep}{0pt}\setlength{\parskip}{0pt}}

\usepackage{mathptmx}

\usepackage{graphics}
\usepackage{amsthm}
\usepackage{amssymb}
\usepackage{bm}

\begin{document}

\def\spacingset#1{\renewcommand{\baselinestretch}%
{#1}\small\normalsize} \spacingset{1}

\title{\bf nflWAR: A Reproducible Method for Offensive Player Evaluation in Football\\\textit{(Extended Edition)}}

  \author{
        Ronald Yurko, Samuel Ventura, and Maksim Horowitz \\
    Department of Statistics \& Data Science, Carnegie Mellon University\\
      }

\maketitle

\bigskip
\begin{abstract}
Unlike other major professional sports, American football lacks comprehensive statistical ratings for player evaluation that are both reproducible and easily interpretable in terms of game outcomes.  Existing methods for player evaluation in football depend heavily on proprietary data, are not reproducible, and lag behind those of other major sports.  We present four contributions to the study of football statistics in order to address these issues.  First, we develop the \texttt{R} package \texttt{nflscrapR} to provide easy access to publicly available play-by-play data from the National Football League (NFL) dating back to 2009.  Second, we introduce a novel multinomial logistic regression approach for estimating the expected points for each play.  Third, we use the expected points as input into a generalized additive model for estimating the win probability for each play.  Fourth, we introduce our \textit{nflWAR} framework, using multilevel models to isolate the contributions of individual offensive skill players, and providing estimates for their individual wins above replacement (\textit{WAR}).  We estimate the uncertainty in each player's \textit{WAR} through a resampling approach specifically designed for football, and we present these results for the 2017 NFL season.  We discuss how our reproducible \textit{WAR} framework, built entirely on publicly available data, can be easily extended to estimate \textit{WAR} for players at any position, provided that researchers have access to data specifying which players are on the field during each play.  Finally, we discuss the potential implications of this work for NFL teams.
\end{abstract}

\noindent%
\vfill

\newpage


\section{Introduction}

Despite the sport's popularity in the United States, public statistical analysis of American football (``football") has lagged behind that of other major sports.  While new statistical research involving player and team evaluation is regularly published in baseball \citep{Albert06, Jensen09, Piette12, Baumer15}, basketball \citep{Kubatko07, Deshpande16}, and hockey \citep{Macdonald11, Gramacy12, Thomas13}, there is limited new research that addresses on-field or player personnel decisions for National Football League (NFL) teams.  Recent work in football addresses topics such as fantasy football \citep{Becker16}, predicting game outcomes \citep{Balreira14}, NFL TV ratings \citep{Grimshaw14}, the effect of ``fan passion'' and league sponsorship on brand recognition \citep{Wakefield12}, and realignment in college football \citep{Jensen14}.  Additionally, with the notable exception of \citet{Lock14}, recent research relating to on-field or player personnel decisions in football is narrowly focused.  For example, \citet{Mulholland14} analyze the success of tight ends in the NFL draft, \citet{Clark13} and \citet{Pasteur14} both provide improved metrics for kicker evaluation, \citet{Martin17} examine the NFL's change in overtime rules, and \citet{Snyder15} focus on discretionary penalties from referees.  Moreover, statistical analysis of football that does tackle on-field or player personnel decisions frequently relies on proprietary and costly data sources, where data quality often depends on potentially biased and publicly unverified human judgment.  This leads to a lack of reproducibility that is well-documented in sports research \citep{Baumer15}. 

In this paper, we posit that (1) objective on-field and player personnel decisions rely on two fundamental categories of statistical analysis in football:  play evaluation and player evaluation, and (2) in order to maintain a standard of objectivity and reproducibility for these two fundamental areas of analysis, researchers must agree on a dataset standard.

\subsection{Previous Work: Evaluating Plays}
\label{sec:prev-work-plays}

The most basic unit of analysis in football is a single play. In order to objectively evaluate on-field decisions and player performance, each play in a football game must be assigned an appropriate value indicating its success or failure. Traditionally, yards gained/lost have been used to evaluate the success of a play. However, this point of view strips away the importance of context in football \citep{Carter71, Carroll88}.  For instance, three yards gained on 3rd and 2 are more valuable than three yards gained on 3rd and 7.  This key point, that not all yards are created equal, has been the foundation for the development of two approaches for evaluating plays:  expected points and win probability.  The expected points framework uses historical data to find the number of points eventually scored by teams in similar situations, while the win probability framework uses historical data to find how often teams in similar situations win the game.  Using these metrics, one can obtain pre-snap and post-snap values of a play (expected points or win probability) and, taking the difference in these values, the value provided by the play itself -- expected points added (EPA) or win probability added (WPA).  These approaches have been recently popularized by Brian Burke's work at \url{www.advancedfootballanalytics.com} and ESPN \citep{Burke_EP, ESPN_total_QBR}.

Most of the best known approaches for calculating expected points do not provide any level of statistical detail describing their methodology.  In most written descriptions, factors such as the down, yards to go for a first down, and field position are taken into account.  However, there is no universal standard for which factors should be considered.  \citet{Carter71} and others essentially use a form of ``nearest neighbors" algorithms \citep{Dasarathy} to identify similar situations based on down, yards to go, and the yard line to then average over the next points scored.  \citet*{Goldner17} describes a Markov model and uses the absorption probabilities for different scoring events (touchdown, field goal, and safety) to arrive at the expected points for a play.  ESPN has a proprietary expected points metric, but does not detail the specifics of how it is calculated \citep{ESPN_EP}. \citet*{Burke_EP} provides an intuitive explanation for what expected points means, but does not go into the details of the calculations.  \citet*{Schatz03} provides a metric called ``defense-adjusted value over average", which is similar to expected points, and also accounts for the strength of the opposing defense.  However, specifics on the modeling techniques are not disclosed.  \citet*{Causey15} takes an exact-neighbors approach, finding all plays with a set of identical characteristics, taking the average outcome, and conducting post-hoc smoothing to calculate expected points.  In this work, Causey explores the uncertainty in estimates of expected points using bootstrap resampling and analyzes the changes in expected point values over time. Causey also provides all code used for this analysis.

Depending on how metrics based on expected points are used, potential problems arise when building an expected points model involving the nature of football games.  The main issue, as pointed out by \citet*{BurkeEP}, involves the score differential in a game.  When a team is leading by a large number of points at the end of a game, they will sacrifice scoring points for letting time run off the clock.  Changes in team behavior in these situations and, more generally, the leverage of a play in terms of its potential effect on winning and losing are not taken into account when computing expected points.  

Analyzing changes in win probability for play evaluation partially resolves these issues.  Compared to expected points models, there is considerably more literature on different methodologies for estimating the win probability of a play in football.  \citet*{Goldner17} uses a Markov model, similar to the approach taken by \citet*{Tango07} in baseball, by including the score differential, time remaining, and timeouts to extend the expected points model.  Burke's approach is primarily empirical estimation by binning plays with adjustments and smoothing.  In some published win probability analyses, random forests have been shown to generate well-calibrated win probability estimates \citep{Causey13, Lock14}.  The approach taken by \citet*{Lock14} also considers the respective strengths of the offensive (possession) and defensive (non-possession) teams.

There are many areas of research that build off of these approaches for valuing plays.  For example, analyses of fourth down attempts and play-calling are very popular \citep{Romer06, Alamar10, Goldner12, Quealy}.  This paper focuses on using play evaluation to subsequently evaluate players, and we discuss prior attempts at player evaluation below.

\subsection{Previous Work:  Evaluating Players}
\label{sec:prev-work-players}

Due to the complex nature of the sport and the limited data available publicly, the NFL lacks comprehensive statistics for evaluating player performance.  While there has been extensive research on situational analysis and play evaluation as described above, there has been considerably less focus on player evaluation.  Existing measures do not accurately reflect a player's value to NFL teams, and they are not interpretable in terms of game outcomes (e.g. points or wins).  Similarly, there are no publicly known attempts for developing a \textit{Wins Above Replacement} (\textit{WAR}) measure for every individual NFL player, as made popular in baseball \citep{Schoenfield12} and other sports \citep{Thomas15}.

Previous methods for player evaluation in football can be broken down into three categories:  within-position statistical comparisons, ad hoc across-position statistical comparisons, and across-position statistical comparisons that rely on proprietary data or human judgment.

\subsubsection{Within-Position Player Evaluation}

Approaches for quantitatively evaluating players who play the same position are numerous, vary by position, and typically lag behind those of other sports.  For comparisons of players at offensive skill positions such as quarterback (QB), running back (RB), wide receiver (WR), and tight end (TE), most analysis relies on basic box score statistics.  These include yards gained via passing, rushing, and/or receiving; touchdowns via passing, rushing, and/or receiving; rushing attempts for RBs; receptions and targets for RBs, WRs, and TEs; completions, attempts, completion percentage, and yard per attempt for QBs; and other similar derivations of simple box score statistics.  These metrics do not account for game situation or leverage. Additionally, they only provide an estimate of a player's relative value to other players at the same position.  We cannot draw meaningful conclusions about cross-positional comparisons.  

Linear combinations of these box score statistics, such as passer rating \citep{Smith73}, are often used to compare players at the same position while taking into account more than just a single box score measure.  Similarly, Pro Football Reference's adjusted net yards per attempt (``ANY/A") expands upon passer rating in that it accounts for sacks and uses a different linear weighting scheme \citep{PFR}.  These metrics involve outdated and/or ad hoc weights, thresholds, and other features.  Passing in the NFL has changed substantially since the conception of the passer rating statistic in 1973, so that the chosen weights and thresholds do not have the same meaning in today's game as they did in 1973.  While ANY/A accounts for sacks and uses a different weighting system, it is hardly a complete measure of QB performance, since it does not account for game situation and leverage.  Perhaps most importantly, both passer rating and ANY/A are not interpretable in terms of game outcomes like points or wins.  

For positions other than QB, RB, WR, and TE, data is limited, since the NFL does not publicly provide information about which players are on the field for a particular play, the offensive and defensive formations (other than the ``shotgun" formation on offense), or the pre- and post-snap locations of players on the field.  For offensive linemen, very little information is available to statistically compare players, as offensive linemen typically only touch the football on broken plays.  For defensive players, the NFL only provides information about which players were directly involved in a play (e.g. the tackler or the defensive back covering a targeted receiver).  As such, with these positions, it is difficult to obtain adequate within-positional comparisons of player value, let alone across-position comparisons.

\subsubsection{Ad Hoc Across-Position Player Evaluation}

Using only box score statistics, it is extremely difficult to ascertain the value of players at different positions.  The fantasy sports industry has attempted to provide across-position estimates of player value using box score statistics.  These estimates typically use ad hoc linear combinations of box score statistics that differ by position, so as to put the in-game statistical performances of players at different positions on comparable scales.  These measures, typically referred to as ``fantasy points", are available for all positions except those on the offensive line.  

Of course, these metrics have several issues.  First, they involve many unjustified or ad hoc weights.  For example, one rushing yard is worth about 40\% of one passing yard in ESPN's standard definitions of these metrics \citep{ESPN_fantasy}, but these relative values are arbitrary.  Second, the definitions are inconsistent, with different on-field events having different values for players of different positions.  For example, defensive interceptions are typically worth three times as much as quarterback interceptions thrown \citep{PFF_fantasy, ESPN_fantasy}.  Third, these measures do not account for context, such as the game situation or the leverage of a given play.  Finally, they are not directly interpretable in terms of game outcomes (e.g. points or wins).

\subsubsection{Player Evaluation with Proprietary Data or Human Judgment}

Outside of the public sphere, there have been irreproducible attempts at within-position statistical comparisons of NFL players.  Pro Football Focus assigns grades to every player in each play, but this approach is solely based on human judgment and proprietary to PFF \citep{Eager17}.  ESPN's total quarterback rating (``QBR") accounts for the situational contexts a QB faces throughout a game \citep{ESPN_total_QBR, Oliver11}.  ESPN uses the following approach when computing QBR:  First, they determine the degree of success or failure for each play.  Second, they divide credit for each play amongst all players involved.  Third, additional adjustments are made for plays of very little consequence to the game outcome. This approach has several important advantages.  In the first step, the EPA is used to assign an objective value to each play.  Another advantage is that some attempt is made to divide credit for a play's success or failure amongst the players involved.  In the approach for NFL player evaluation we propose in this paper, we loosely follow these same two steps.

ESPN's QBR has some disadvantages, however.  First and most importantly, Total QBR is not directly reproducible, since it relies on human judgment when evaluating plays.  ``The details of every play (air yards, drops, pressures, etc.) are charted by a team of trained analysts in the ESPN Stats \& Information Group. Every play of every game is tracked by at least two different analysts to provide the most accurate representation of how each play occurred" \citep{ESPN_total_QBR}.  Additionally, while QBR down-weights plays in low-leverage situations, the approach for doing so is not clearly described and appears to be ad hoc.  Finally, QBR is limited only to the QB position.

The only public approach for evaluating players at all positions according to common scale is Pro Football Reference's ``approximate value" (AV) statistic \citep{Drinen}.  Using a combination of objective and subjective analysis, AV attempts to assign a single numerical value to a player's performance in any season since 1950, regardless of the player's position.  AV has some subjective components, such as whether or not a lineman was named to the NFL's ``all-pro" team, and whether a running back reaches the arbitrary threshold of 200 carries.  Additionally, since AV uses linear combinations of end-of-season box score statistics to evaluate players, it does not take into account game situation, opponent, or many other contextual factors that may play a role in the accumulation of box score statistics over the course of a season.  Finally, although the basis of many AV calculations involves points scored and allowed, AV is not interpretable in terms of game outcomes.

\subsection{Our Framework for Evaluating NFL Plays and Players}

In order to properly evaluate players, we need to allocate a portion of a play's value to each player involved \citep{ESPN_total_QBR}. \citet*{Baumer17} details the long history of division of credit modeling as a primary driver of research in sports analytics, with origins in evaluating run contributions in baseball.  However, in comparison to baseball, every football play is more complex and interdependent, with the 22 players on the field contributing in many ways and to varying degrees.  A running play depends not only on the running back but the blocking by the linemen, the quarterback's handoff, the defensive matchup, the play call, etc.  A natural approach is to use a regression-based method, with indicators for each player on the field for a play, providing an estimate of their marginal effect.  This type of modeling has become common in basketball and hockey, because it accounts for factors such as quality of teammates and competition \citep{Rosenbaum04, Kubatko07, Macdonald11, Gramacy12, Thomas13}.

We present four contributions to the study of football statistics in order to address the issues pertaining to play evaluation and player evaluation outlined above:

\begin{enumerate}
  \item The \texttt{R} package \texttt{nflscrapR} to provide easy access to publicly available NFL play-by-play data (Section \ref{sec:data}).
  \item A novel approach for estimating expected points using a multinomial logistic regression model, which more appropriately models the ``next score" response variable (Section \ref{sec:ep}).
  \item A generalized additive model for estimating the win probability using the expected points as input (Section \ref{sec:wp}).
  \item Our \textit{nflWAR} framework, using multilevel models to isolate offensive skill player contribution and estimate their \textit{WAR} (Section \ref{sec:nflwar}).  
\end{enumerate}

We use a sampling procedure similar to \citet{Baumer15} to estimate uncertainty in each player's seasonal \textit{WAR}.  Due to the limitations of publicly available data, the primary focus of this paper is on offensive skill position players: QB, RB, WR, and TE.  However, we present a novel metric that serves as a proxy for measuring a team's offensive line performance on rushing plays.  Furthermore, the reproducible framework we introduce in this paper can also be easily extended to estimate \textit{WAR} for all positions given the appropriate data.  Researchers with data detailing which players are on the field for every play can use the framework provided in Section \ref{sec:road_to_war} to estimate \textit{WAR} for players at all positions.

Our \textit{WAR} framework has several key advantages.  First, it is fully reproducible:  it is built using only public data, with all code provided and all data accessible to the public. Second, our expected points and win probability models are well-calibrated and more appropriate from a statistical perspective than other approaches. Third, player evaluation with \textit{WAR} is easily interpretable in terms of game outcomes, unlike prior approaches to player evaluation in the NFL discussed above.  The replacement level baseline informs us how many wins a player adds over a readily available player.  This is more desirable than comparing to average from the viewpoint of an NFL front office, as league average performance is still valuable in context \citep{Baumer15}.  Fourth, the multilevel model framework accounts for quality of teammates and competition.  Fifth, although this paper presents \textit{WAR} using our expected points and win probability models for play evaluation, researchers can freely substitute their own approaches for play evaluation without any changes to the framework for estimating player \textit{WAR}. Finally, we recognize the limitations of point estimates for player evaluation and provide estimates of the uncertainty in a player's \textit{WAR}.

\section{Play-by-Play Data with \texttt{nflscrapR}}
\label{sec:data}

Data in professional sports comes in many different forms. At the
season-level, player and team statistics are typically available dating
back to the 1800s \citep[\citet{Phillips}]{Lahman}. At the game-level,
player and team statistics have been tracked to varying degrees of
detail dating back several decades \citep{Lahman}. Within games, data is
available to varying degrees of granularity across sports and leagues.
For example, Major League Baseball (MLB) has play-by-play data at the
plate appearance level available dating back several decades
\citep{Lahman}, while the National Hockey League (NHL) only began
releasing play-by-play via their real-time scoring system in the 2005-06
season \citep{Thomas17}.

Play-by-play data, or information specifying the conditions, features,
and results of an individual play, serves as the basis for most modern
sports analysis in the public sphere \citep[\citet{Macdonald11},
\citet{Lock14}, \citet{Thomas17}]{Kubatko07}. Outside of the public
sphere, many professional sports teams and leagues have access to data
at even finer levels of granularity, e.g.~via optical player tracking
systems in the National Basketball Association, MLB, and the English
Premier League that track the spatial distribution of players and
objects at multiple times per second. The NFL in 2016 began using
radio-frequency identification (RFID) technology to track the locations
of players and the football \citep{CBS}, but as of mid 2018, this data
is not available publicly, and NFL teams have only just gained accessed to the data beyond their own players. In almost all major professional sports leagues, play-by-play data is provided and includes information on in-game
events, players involved, and (usually) which players are actively
participating in the game for each event
\citep[\citet{Lahman}]{Thomas17}.

Importantly, this is not the case for the NFL. While play-by-play data
is available through the NFL.com application programming interface
(API), the league does not provide information about which players are
present on the playing field for each play, what formations are being
used (aside from the ``shotgun'' formation), player locations, or
pre-snap player movement. This is extremely important, as it limits the
set of players for which we can provide estimates of their contribution
to game outcomes (e.g.~points scored, points allowed, wins, losses,
etc).

We develop an \texttt{R} package \citep{R17}, called \texttt{nflscrapR},
that provides users with clean datasets, box score statistics, and more
advanced metrics describing every NFL play since 2009
\citep{Horowitz17}. This package was inspired largely by other
\texttt{R} packages facilitating the access of sports data. For hockey,
\texttt{nhlscrapR} provides clean datasets and advanced metrics to use
for analysis for NHL fans \citep{Thomas17}. In baseball, the R packages
\texttt{pitchRx} \citep{Sievert15}, \texttt{Lahman} \citep{Lahman}, and
\texttt{openWAR} \citep{Baumer15} provide tools for collecting MLB data
on the pitch-by-pitch level and building out advanced player evaluation
metrics. In basketball, \texttt{ballR} \citep{Elmore17} provides
functions for collecting data from \texttt{basketball-reference.com}.

Each NFL game since 2009 has a 10 digit game identification number (ID)
and an associated set of webpages that includes information on the scoring
events, play-by-play, game results, and other game data. The API
structures its data using JavaScript Object Notation (JSON) into three
major groups: game outcomes, player statistics at the game level, and
play-by-play information. The design of the \texttt{nflscrapR} package
closely mimics the structure of the JSON data in the API, with four main
functions described below:

\texttt{season\_games()}: Using the data structure outputting end of
game scores and team matchups, this function provides end of game
results with an associated game ID and the home and away teams
abbreviations.

\texttt{player\_game()}: Accessing the player statistics object in the
API's JSON data, this function parses the player level game summary data
and creates a box-score-like data frame. Additional functions provide
aggregation functionality:\\ \texttt{season\_player\_game()} binds the
results of \texttt{player\_game()} for all games in a season, and
\texttt{agg\_player\_season()} outputs a single row for each player with
their season total statistics.

\texttt{game\_play\_by\_play()}: This is the most important function in
\texttt{nflscrapR}. The function parses the listed play-by-play data
then uses advanced regular expressions and other data manipulation tasks
to extract detailed information about each play (e.g.~players involved
in action, play type, penalty information, air yards gained, yards
gained after the catch, etc.). The \texttt{season\_play\_by\_play()}
binds the results of \texttt{game\_play\_by\_play()} for all games in a
season.

\texttt{season\_rosters()}: This function outputs all of the rostered
players on a specified team in a specified season and includes their
name, position, unique player ID, and other information.

For visualization purposes we also made a dataset, \texttt{nflteams}
available in the package which includes the full name of all 32 NFL
teams, their team abbreviations, and their primary
colors\footnote{Some of this information is provided through Ben Baumer's \texttt{R} package \texttt{teamcolors} \citep{Baumer17b}}.

In addition to the functions provided in \texttt{nflscrapR}, we provide
downloadable versions in comma-separated-value format, along with a
complete and frequently updating data dictionary, at
\texttt{https://github.com/ryurko/nflscrapR-data}. The datasets provided on this website included play-by-play from 2009 -- 2017, game-by-game player level statistics, player-season total statistics, and team-season total statistics. These datasets are made available to allow users familiar with other software to do research in the realm of football analytics. Table \ref{table-pbp} gives a brief overview of some of the more important variables used for evaluating plays in Section
\ref{sec:ep_wp_model}.

\begin{table}
\centering
\caption{Description of the play-by-play dataset.}
\label{table-pbp}
\begin{tabular}{p{3cm} p{9cm}}
\hline \\ [-1.5ex]
Variable & Description \\ [1ex]
\hline \\ [-1.5ex]
Possession Team & Team with the ball on offense (opposing team is on defense) \\ [1ex]
Down & Four downs to advance the ball ten (or more) yards \\ [1ex]
Yards to go & Distance in yards to advance and convert first down \\ [1ex]
Yard line & Distance in yards away from opponent's endzone (100 to zero)  \\ [1ex]
Time Remaining & Seconds remaining in game, each game is 3600 seconds long (four quarters, halftime, and a potential overtime) \\ [1ex]
Score differential & Difference in score between the possession team and opposition \\
\hline 
\end{tabular}
\end{table}

\section{Evaluating Plays with Expected Points and Win Probability}
\label{sec:ep_wp_model}

As described in Section \ref{sec:prev-work-plays}, expected points and
win probability are two common approaches for evaluating plays. These
approaches have several key advantages: They can be calculated using
only data provided by the NFL and available publicly, they provide
estimates of a play's value in terms of real game outcomes (i.e.~points
and wins), and, as a result, they are easy to understand for both
experts and non-experts.

Below, we introduce our own novel approaches for estimating expected
points ($EP$) and win probability ($WP$) using publicly available
data via \texttt{nflscrapR}.

\subsection{Expected Points}
\label{sec:ep}

While most authors take the average ``next score'' outcome of similar
plays in order to arrive at an estimate of $EP$, we recognize that
certain scoring events become more or less likely in different
situations. As such, we propose modeling the probability for each of
the scoring events directly, as this more appropriately accounts for the
differing relationships between the covariates in Table \ref{table-pbp}
and the different categories of the ``next score'' response. Once we
have the probabilities of each scoring event, we can trivially estimate
expected points.

\subsubsection{Multinomial Logistic Regression}

To estimate the probabilities of each possible scoring event conditional
on the current game situation, we use multinomial logistic regression.
For each play, we find the next scoring event within the same half (with
respect to the possession team) as one of the seven possible events:
touchdown (7 points), field goal (3 points), safety (2 points), no score
(0 points), opponent safety (-2 points), opponent field goal (-3
points), and opponent touchdown (-7 points). Here, we ignore point after
touchdown (PAT) attempts, and we treat PATs separately in Section
\ref{sec:pat_fg}.

\begin{figure}[!h]
\includegraphics[width=14cm]{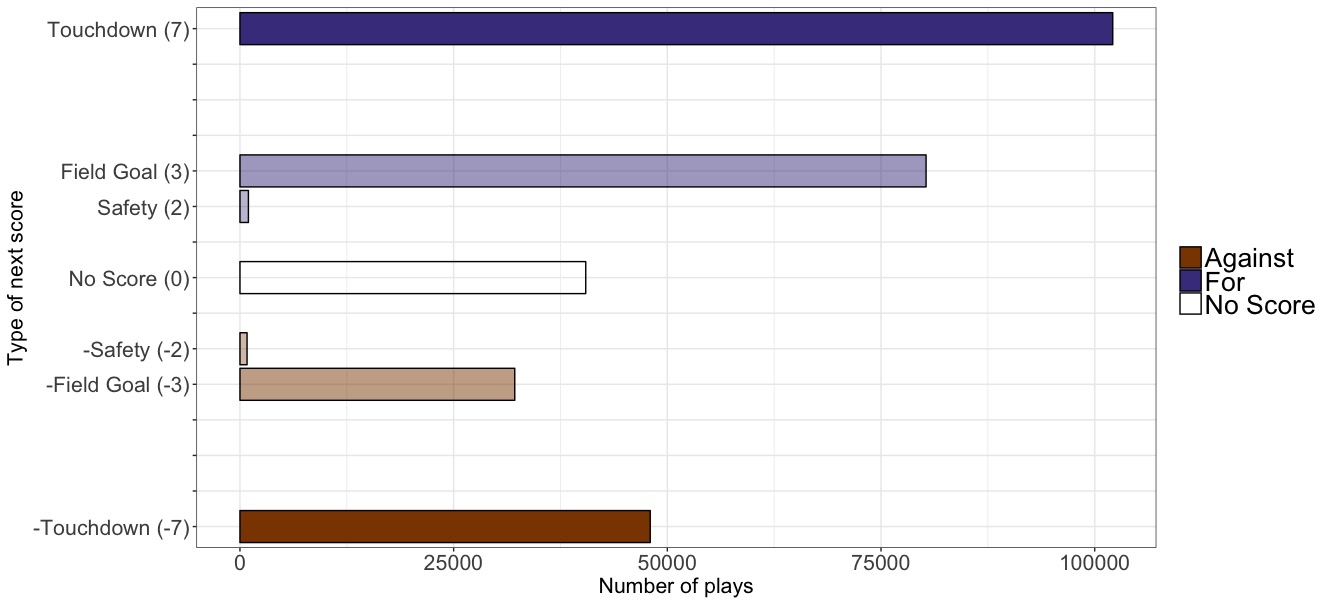}
\centering
\caption{Distribution of next scoring events for all plays from 2009-16, with respect to the possession team.}
\label{next-score-bar}
\end{figure}

\autoref{next-score-bar} displays the distribution of the different type
of scoring events using data from NFL regular season games between 2009
and 2016, with each event located on the y-axis based on their
associated point value $y$. This data consists of 304,896 non-PAT
plays, excluding QB kneels (which are solely used to run out the clock
and are thus assigned an $EP$ value of zero). The gaps along the
y-axis between the different scoring events reinforce our decision to
treat this as a classification problem rather than modeling the point
values with linear regression -- residuals in such a model will not meet
the assumptions of normality. While we use seven points for a touchdown
for simplicity here, our multinomial logistic regression model generates
the probabilities for the events agnostic of the point value. This is
beneficial, since it allows us to flexibly handle PATs and two-point
attempts separately. We can easily adjust the point values
associated with touchdowns to reflect changes in the league's scoring
environment.

\begin{table}
\centering
\caption{Description of variables for the $EP$ model.}
\label{table-ep-vars}
\begin{tabular}{ p{3cm} p{9cm}}
\hline \\ [-1.5ex]
Variable & Variable description \\ [1ex]
\hline \\ [-1.5ex]
Down & The current down (1st, 2nd, 3rd, or 4th\\ [1ex]
Seconds & Number of seconds remaining in half \\ [1ex]
Yardline & Yards from endzone (0 to 100)  \\ [1ex]
log(YTG) & Log transformation of yards to go for a first down \\ [1ex]
GTG & Indicator for whether or not it is a goal down situation \\ [1ex]
UTM & Indicator for whether or not time remaining in the half is under two minutes \\ [1ex]
\hline
\end{tabular}
\end{table}

We denote the covariates describing the game situation for each play as
$\mathbf{X}$, which are presented in Table \ref{table-ep-vars}, and the
response variable:
\begin{align}
\label{ep-response}
Y \in & \{\textrm{Touchdown}\ (\textbf{7}),\ \textrm{Field Goal}\ (\textbf{3}),\ \textrm{Safety}\ (\textbf{2}),\ \textrm{No Score}\ (\textbf{0}), \nonumber \\ 
&-\textrm{Touchdown}\ (\textbf{-7}),\ -\textrm{Field Goal}\ (\textbf{-3}),\ -\textrm{Safety}\ (\textbf{-2})\}
\end{align}

The model is specified with six logit transformations relative to the ``No Score''
event with the following form: 
\begin{align}
  \text{log}(\frac{P(Y=Touchdown|\mathbf{X})}{P(Y=No\ Score|\mathbf{X})}) & = \mathbf{X}\cdot \boldsymbol{\beta}_{Touchdown}, \nonumber \\
  \text{log}(\frac{P(Y=Field\ Goal|\mathbf{X})}{P(Y=No\ Score|\mathbf{X})}) & = \mathbf{X}\cdot \boldsymbol{\beta}_{Field\ Goal}, \nonumber \\
  \vdots  \\
  \text{log}(\frac{P(Y=-Touchdown|\mathbf{X})}{P(Y=No\ Score|\mathbf{X})}) & = \mathbf{X}\cdot \boldsymbol{\beta}_{-Touchdown}, \nonumber
\end{align}

\noindent where $\boldsymbol{\beta}_y$ is the corresponding coefficient vector for the type of next scoring event. Using the generated probabilities for each of the possible scoring
events, $P(Y = y|\mathbf{X})$, we simply calculate the expected
points ($EP$) for a play by multiplying each event's predicted
probability with its associated point value \(y\):
\begin{equation}
\label{ep-formula}
EP = E[Y | \mathbf{X}] = \sum_y y \cdot P(Y=y | \mathbf{X}).
\end{equation}

\hypertarget{observation-weighting}{%
\subsubsection{Observation Weighting}\label{observation-weighting}}

\begin{figure}[!h]
\includegraphics[width=16cm]{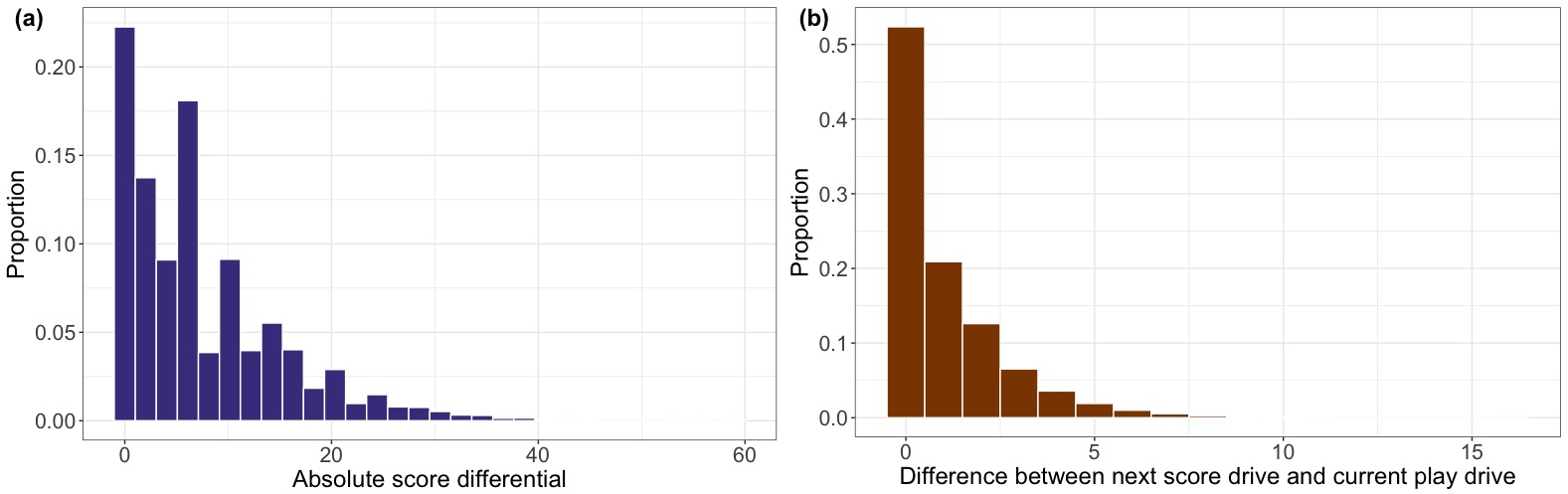}
\centering
\caption{Distributions for (a) absolute score differential and (b) number of drives until next score (excluding plays without a next score event).}
\label{score_diff_hist}
\end{figure}

Potential problems arise when building an expected points model
because of the nature of football games. The first issue, as pointed out
by \citet{BurkeEP}, regards the score differential in a game. When a
team is leading by a large number of points at the end of a game they
will sacrifice scoring points for letting time run off the clock. This
means that plays with large score differentials can exhibit a different
kind of relationship with the next points scored than plays with tight
score differentials. Although others such as Burke only use the subset
of plays in the first and third quarter where the score differential is
within ten points, we don't exclude any observations but instead use a
weighting approach. \autoref{score_diff_hist}(a) displays the distribution
for the absolute score differential, which is clearly skewed right, with a higher proportion of plays possessing smaller score
differentials. Each play \(i \in \{1, \hdots, n\}\), in the modeling data of regular season games from 2009 to 2016, is assigned a weight
\(w_i\) based on the score differential \(S\) scaled from zero to one
with the following function:
\begin{equation}
\label{weight}
w_i = w(S_i) = \frac{\underset{i}{max}(|S_i|) - |S_i|}{\underset{i}{max}(|S_i|) - \underset{i}{min}(|S_i|)}.
\end{equation}

In addition to score differential, we also weight plays according to
their ``distance'' to the next score in terms of the number of drives.
For each play \(i\), we find the difference in the number of drives from
the next score \(D\): \(D_i = d_{next\  score} - d_i\), where
\(d_{next\  score}\) and \(d_i\) are the drive numbers for the next
score and play \(i\), respectively. For plays in the first half, we
stipulate that \(D_i = 0\) if the \(d_{next\  score}\) occurs in the
second half, and similarly for second half plays for which the next
score is in overtime. \autoref{score_diff_hist}(b) displays the
distribution of \(D_i\) excluding plays with the next score as ``No
Score.'' This difference is then scaled from zero to one in the same way
as the score differential in \autoref{weight}. The score differential
and drive score difference weights are then added together and again
rescaled from zero to one in the same manner resulting in a combined
weighting scheme. By combining the two weights, we are placing equal
emphasis on both the score differential and the number of drives until
the next score and leave adjusting this balance for future work.

\hypertarget{model-selection-with-calibration}{%
\subsubsection{Model Selection with
Calibration}
\label{model-selection-with-calibration}}

Since our expected points model uses the probabilities for each scoring
event from multinomial logistic regression, the variables and
interactions selected for the model are determined via calibration
testing, similar to the criteria for evaluating the win probability
model in \citet{Lock14}. The estimated probability for each of the seven
scoring events is binned in five percent increments (20 total possible
bins), with the observed proportion of the event found in each bin. If
the actual proportion of the event is similar to the bin's estimated
probability then the model is well-calibrated. Because we are generating
probabilities for seven events, we want a model that is well-calibrated
across all seven events. To objectively compare different models, we
first calculate for scoring event \(y\) in bin \(b \in \{1,\hdots, B\}\)
its associated error \(e_{y,b}\):
\begin{equation}
e_{y,b} = |\hat{P_b}(Y=y) - P_b(Y=y)|,
\end{equation}

\noindent where \(\hat{P_b}(Y=y)\) and \(P_b(Y=y)\) are the predicted
and observed probabilities, respectively, in bin \(b\). Then, the
overall calibration error \(e_y\) for scoring event \(y\) is found by
averaging \(e_{y,b}\) over all bins, weighted by the number of plays in
each bin, \(n_{y,b}\):
\begin{equation}
e_y = \frac{1}{n_y}\sum_b n_{y,b} \cdot e_{y,b},
\end{equation}

\noindent where \(n_y = \sum_b n_{y,b}\). This leads to the model's
calibration error \(e\) as the average of the seven \(e_y\) values,
weighted by the number of plays with scoring event \(y\), \(n_{y}\):
\begin{equation}
e = \frac{1}{n}\sum_y n_y \cdot e_y,
\end{equation}

\noindent where \(n = \sum_y n_y\), the number of total plays. This
provides us with a single statistic with which to evaluate models, in
addition to the calibration charts.

\begin{figure}[!h]
\includegraphics[width=16cm]{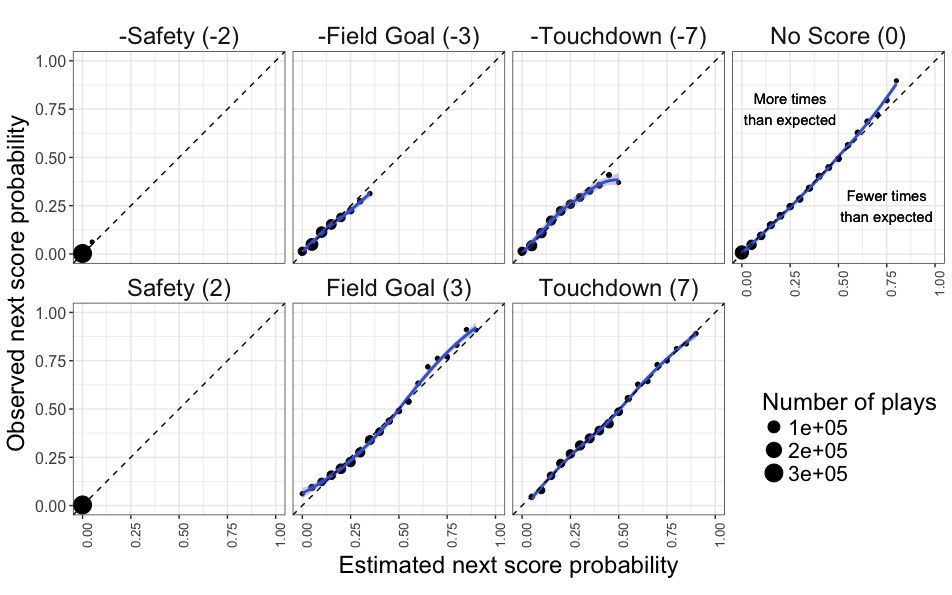}
\centering
\caption{Expected points model LOSO CV calibration results by scoring event.}
\label{ep_cal}
\end{figure}

We calculate the model calibration error using leave-one-season-out
cross-validation (LOSO CV) to reflect how the \texttt{nflscrapR} package
will generate the probabilities for plays in a season it has not yet
observed. The model yielding the best LOSO CV calibration results uses
the variables presented in \autoref{table-ep-vars}, along with three
interactions: \(\text{log}(\mbox{YTG})\) and Down, Yardline and Down, and
\(\text{log}(\mbox{YTG})\) and GTG. \autoref{ep_cal} displays the selected
model's LOSO CV calibration results for each of the seven scoring
events, resulting in \(e \approx 0.013\). The dashed lines along the
diagonal represent a perfect fit, i.e.~the closer to the diagonal points
are the more calibrated the model. Although time remaining is typically
reserved for win probability models \citep{Goldner17}, including the
seconds remaining in the half, as well as the indicator for under two
minutes, improved the model's calibration, particularly with regards to
the ``No Score'' event. We also explored the use of an ordinal logistic regression model which assumes equivalent effects as the scoring value increases, but found the LOSO CV calibration results to be noticeably worse with \(e \approx 0.022\).

\hypertarget{pats-and-field-goals}{%
\subsubsection{PATs and Field Goals}\label{pats-and-field-goals}}

\label{sec:pat_fg}

As noted earlier, we treat PATs (extra point attempts and two-point
attempts) separately. For two-point attempts, we simply use the
historical success rate of 47.35\% from 2009-2016, resulting in
\(EP = 2 \cdot 0.4735 = 0.9470\). Extra point attempts use the
probability of successfully making the kick from a generalized additive
model (see Section \ref{gam}) that predicts the probability of making the
kick, \(P(M)\) for both extra point attempts and field goals as a smooth
function of the kick's distance, \(k\) (total of 16,906 extra point and
field goal attempts from 2009-2016):
\begin{equation}
\text{log}(\frac{P(M)}{1-P(M)}) = s(k).
\end{equation}

The expected points for extra point attempts is this predicted
probability of making the kick, since the actual point value of a PAT is
one. For field goal attempts, we incorporate this predicted probability
of making the field goal taking into consideration the cost
of missing the field goal and turning the ball over to the opposing
team. This results in the following override for field goal attempts:
\begin{equation}
EP_{field\ goal\ attempt} = P(M)\cdot 3 + (1 - P(M))\cdot (-1)\cdot E[Y|X=m],
\end{equation}

\noindent where \(E[Y|X=m]\) is the expected points from the multinomial
logistic regression model but assuming the opposing team has taken
possession from a missed field goal, with the necessary adjustments to
field position and time remaining (eight yards and 5.07 seconds,
respectively, estimated from NFL regular season games from 2009 to 2016), and multiplying by
negative one to reflect the expected points for the team attempting the
field goal. Although these calculations are necessary for proper
calculation of the play values $\delta_{f,i}$ discussed in Section
\ref{epa-wpa}, we note that this is a rudimentary field goal model only
taking distance into account. Enhancements could be made with additional
data (e.g.~weather data, which is not made available by the NFL) or by
using a model similar to that of \citet{Morris15}, but these are beyond
the scope of this paper.

\subsubsection{Expected Points by Down and Yard Line}

\begin{figure}[!h]
\includegraphics[width=16cm]{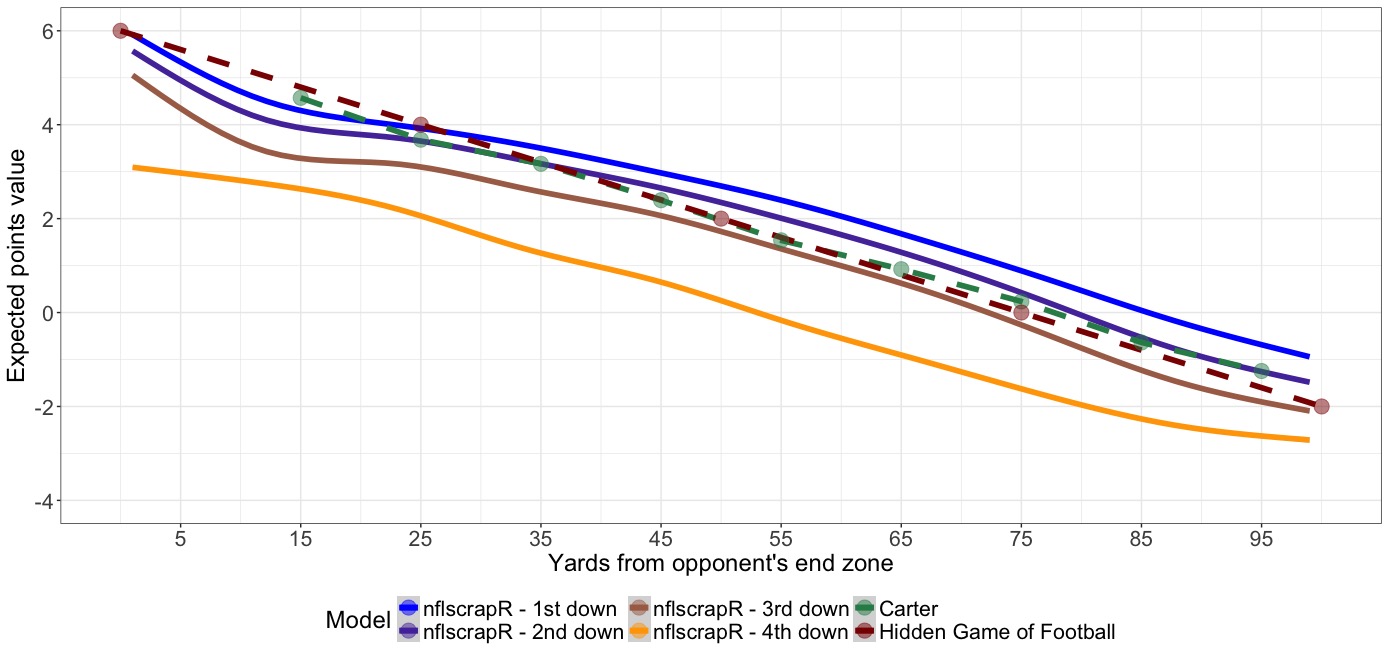}
\centering
\caption{Comparison of historical models and \texttt{nflscrapR} expected points value, based on distance from opponent's end zone by down.}
\label{ep_comp}
\end{figure}

\begin{figure}[!h]
\includegraphics[width=16cm]{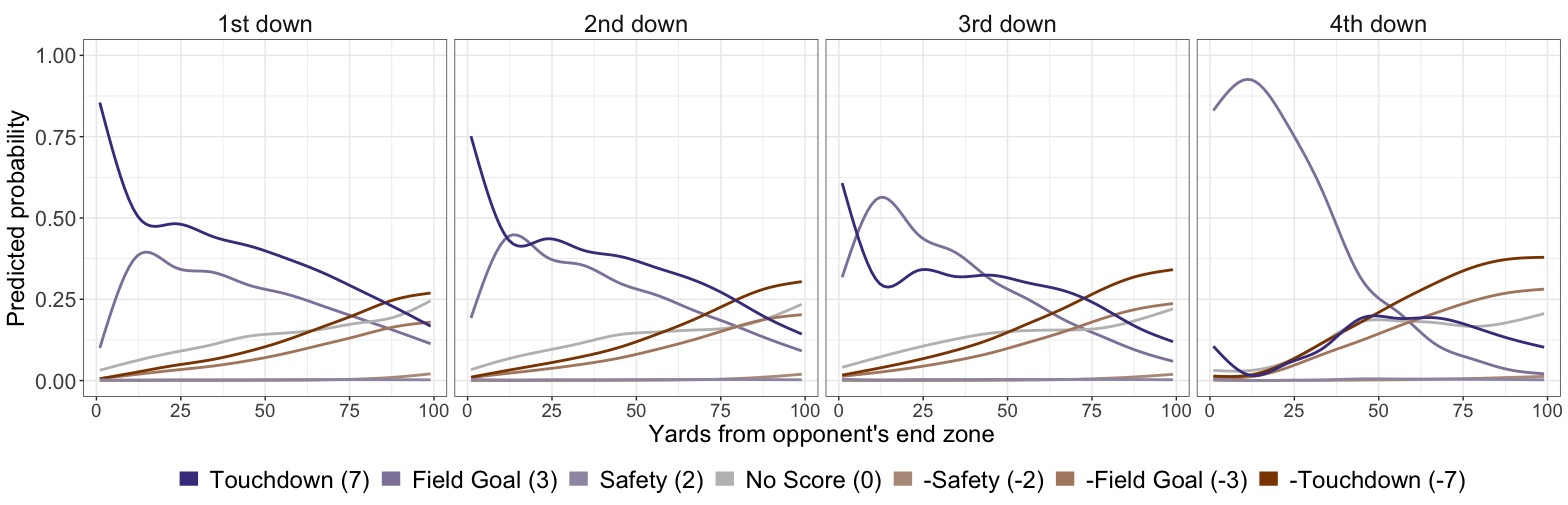}
\centering
\caption{Relationship between next score event probabilities and field position by down.}
\label{ep_probs_chart}
\end{figure}

For reference, \autoref{ep_comp} displays the relationship between the
field position and the $EP$ for our multinomial logistic regression
model available via \texttt{nflscrapR} compared to the previous
relationships found by \citet{Carter71} and \citet{Carroll88}. We
separate the \texttt{nflscrapR} model by down to show its importance,
and in particular the noticeable drop for fourth down plays and how they
exhibit a different relationship near the opponent's end zone as
compared to other downs. To provide context for what is driving the
difference, \autoref{ep_probs_chart} displays the relationship between
each of the next score probabilities and field position by down. Clearly
on fourth down, the probability of a field goal attempt overwhelms the
other possible events once within 50 yards of the opponent's end zone.

\subsection{Win Probability}
\label{sec:wp}

Because our primary focus in this paper is in player evaluation, we
model win probability without taking into account the teams playing (i.e.~we do not
include indicators for team strength in the win probability model). As a
result, every game starts with each team having a 50\% chance of
winning. Including indicators for a team's overall, offensive, and/or
defensive strengths would artificially inflate (deflate) the
contributions made by players on bad (good) teams in the models
described in Section \ref{sec:nflwar}, since their team's win
probability would start lower (higher).

Our approach for estimating \(WP\) also differs from the others
mentioned in Section \ref{sec:prev-work-plays} in that we incorporate the
estimated \(EP\) directly into the model by calculating the expected
score differential for a play. Our expected points model already
produces estimates for the value of the field position, yards to go, etc
without considering which half of the game or score. When including the
variables presented in Table \ref{table-wp-vars}, we arrive at a
well-calibrated \(WP\) model.

\begin{table}
\centering
\caption{Description of selected variables for the win probability model.  Note:  $S$ is the score differential at the current play.}
\label{table-wp-vars}
\begin{tabular}{ p{3cm} p{9cm}}
\hline \\ [-1.5ex]
Variable & Variable description \\ [1ex]
\hline \\ [-1.5ex]
$E[S]$ & Expected score differential = $EP + S$ \\ [1ex]
$s_{g}$ & Number of seconds remaining in game \\ [1ex]
$E[\frac{S}{s_{g} +1}]$ & Expected score time ratio\\ [1ex]
$h$ & Current half of the game (1st, 2nd, or overtime)  \\ [1ex]
$s_h$ & Number of seconds remaining in half \\ [1ex]
$u$ & Indicator for whether or not time remaining in half is under two minutes \\ [1ex]
$t_{off}$ & Time outs remaining for offensive (possession) team \\ [1ex]
$t_{def}$ & Time outs remaining for defensive team \\ [1ex]
\hline
\end{tabular}
\end{table}

\subsubsection{Generalized Additive Model}
\label{gam}

We use a generalized additive model (GAM) to estimate the possession
team's probability of winning the game conditional on the current game
situation. GAMs have several key benefits that make them ideal for
modeling win probability: They allow the relationship between the
explanatory and response variables to vary according to smooth,
non-linear functions. They also allow for linear relationships and can
estimate (both ordered and unordered) factor levels. We find that this
flexible, semi-parametric approach allows us to capture nonlinear
relationships while maintaining the many advantages of using linear
models. Using a logit link function, our \(WP\) model takes the form:
\begin{equation}
\text{log}(\frac{P(\mbox{Win})}{P(\mbox{Loss})}) = s(E[S]) + s(s_h)\cdot h + s(E[\frac{S}{s_{g} +1}]) + h \cdot u \cdot t_{off} + h\cdot u \cdot t_{def},
\end{equation}

\noindent where \(s\) is a smooth function while \(h\), \(u\),
\(t_{off}\), and \(t_{def}\) are linear parametric terms defined in \autoref{table-wp-vars}. By taking the
inverse of the logit we arrive at a play's \(WP\).

\hypertarget{win-probability-calibration}{%
\subsubsection{Win Probability
Calibration}\label{win-probability-calibration}}

\begin{figure}[!h]
\includegraphics[width=16cm]{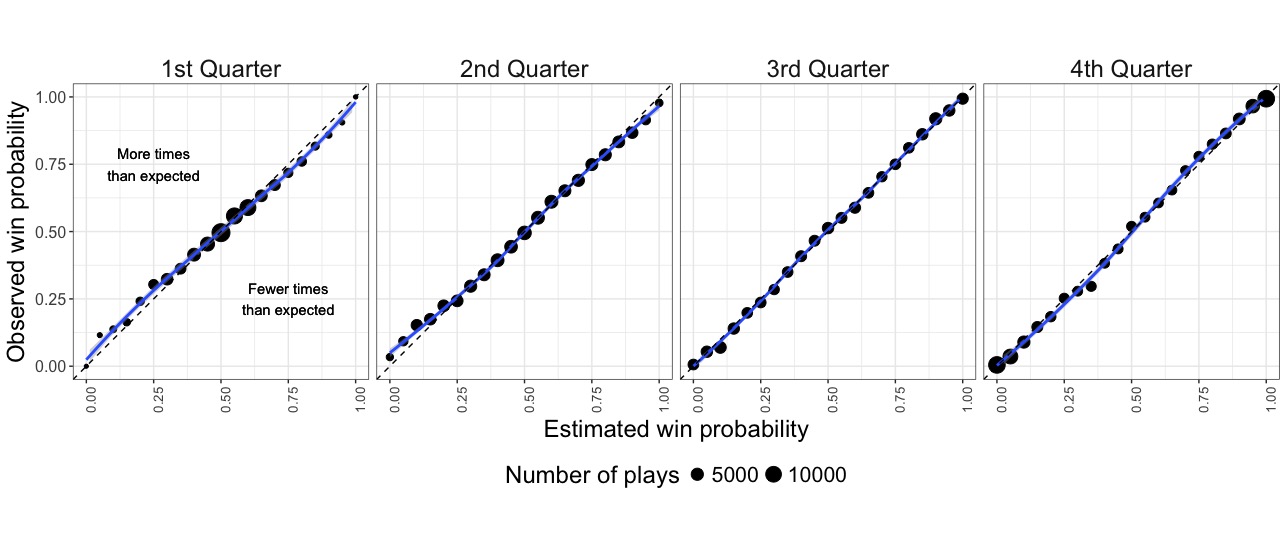}
\centering
\caption{Win probability model LOSO CV calibration results by quarter.}
\label{wp_cal}
\end{figure}

Similar to the evaluation of the \(EP\) model, we again use LOSO CV to
select the above model, which yields the best calibration results.
\autoref{wp_cal} shows the calibration plots by quarter, mimicking the
approach of \citet{Lopez17} and \citet{Yam18}, who evaluate both our
\(WP\) model and that of \citet{Lock14}. The observed proportion of wins
closely matches the expected proportion of wins within each bin for each
quarter, indicating that the model is well-calibrated across all
quarters of play and across the spectrum of possible win probabilities.
These findings match those of \citet{Yam18}, who find ``no obvious
systematic patterns that would signal a flaw in either model.''

\subsubsection{Win Probability Example}

\begin{figure}[!h]
\includegraphics[width=16cm]{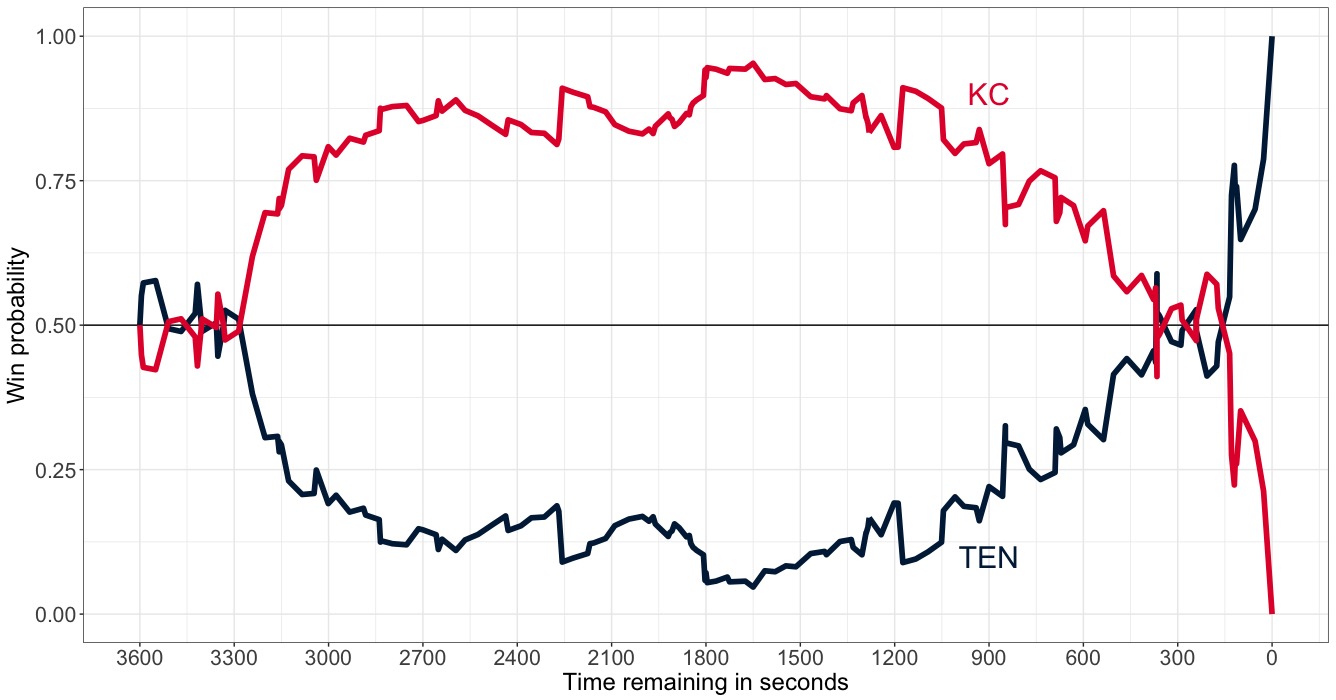}
\centering
\caption{Win probability chart for 2017 AFC Wild Card game.}
\label{wp_example}
\end{figure}

An example of a single game \(WP\) chart is provided in
\autoref{wp_example} for the 2017 American Football Conference (AFC)
Wild Card game between the Tennessee Titans and Kansas City Chiefs.  The game starts with both teams having an equal chance of winning, with minor variations until the score differential changes (in this case, in favor of Kansas City). Kansas City led 21-3 after the first half, reaching a peak win probability of roughly 95\% early in the third quarter, before giving up 19 unanswered points in the second half and losing to Tennessee 22-21.

\subsection{Expected Points Added and Win Probability Added}
\label{epa-wpa}

In order to arrive at a comprehensive measure of player performance, each
play in a football game must be assigned an appropriate value
\(\delta_{f,i}\) that can be represented as the change from state \(i\)
to state \(f\):
\begin{equation}
\label{play-value}
\delta_{f,i} = \boldsymbol{V}_{f} - \boldsymbol{V}_i,
\end{equation}

\noindent where \(\boldsymbol{V}_{f}\) and \(\boldsymbol{V}_{i}\) are
the associated values for the ending and starting states respectively.
We represent these values by either a play \(i\)'s expected points
(\(EP_i\)) or win probability (\(WP_i\)).

Plugging our \(EP\) and \(WP\) estimates for the start of play \(i\) and
the start of the following play \(f\) into \autoref{play-value}'s values
for \(\boldsymbol{V}_i\) and \(\boldsymbol{V}_f\) respectively provides
us with the two types of play valuations \(\delta_{f,i}\): (1) the
change in point value as expected points added (\(EPA\)), and (2) the
change in win probability as win probability added (\(WPA\)). For
scoring plays, we use the associated scoring event's value \(y\) as
\(\boldsymbol{V}_f\) in place of the following play's \(EP\) to reflect
that the play's value is just connected to the difference between the
scoring event and the initial state of the play. As an example, during Super Bowl LII the Philadelphia Eagles' Nick Foles received a touchdown when facing fourth down on their opponent's one yard line with thirty-eight seconds remaining in the half. At the start of the play the Eagles' expected points was \(\boldsymbol{V}_{i}\ \approx 2.78\), thus resulting in $EPA \approx 7 - 2.78 = 4.22$. In an analogous calculation, this famous play known as the ``Philly special" resulted in $WPA \approx 0.1266$ as the Eagles' increased their lead before the end of the half.

For passing plays, we can additionally take advantage of \emph{air
yards} (perpendicular distance in yards from the line of scrimmage to
the yard line at which the receiver was targeted or caught the ball) and
\emph{yards after catch} (perpendicular distance in yards from the yard
line at which the receiver caught the ball to the yard line at which the
play ended), for every passing play available with \texttt{nflscrapR}.
Using these two pieces, we can determine the hypothetical field position
and whether or not a turnover on downs occurs to separate the value of a
play from the air yards versus the yards after catch. For each completed
passing play, we break the estimation of \(EP\) and \(WP\)
into two plays -- one comprising everything leading up to the catch, and
one for the yards after the catch. Because the models rely on the
seconds remaining in the game, we make an adjustment to the time
remaining by subtracting the average length of time for incomplete
passing plays, 5.7
seconds\footnote{This estimate could be improved in future work if information about the time between the snap and the pass becomes available.}.
We then use the \(EP\) or \(WP\) through the air as \(\boldsymbol{V}_f\)
in \autoref{play-value} to estimate \(EPA_{i,air}\) or \(WPA_{i,air}\),
denoting these as \(\delta_{f,i,air}\). We estimate the value of yards
after catch, \(\delta_{f,i,yac}\), by taking the difference between the
value of the following play \(\boldsymbol{V}_f\) and the value of the
air yards, \(\delta_{f,i,air}\). We use this approach to calculate both
\(EPA_{i,yac}\) and \(WPA_{i,yac}\).

\section{Evaluating Players with nflWAR}
\label{sec:nflwar}

We use the play values calculated in Section
\ref{sec:ep_wp_model} as the basis for a statistical estimate of wins
above replacement (\textit{WAR}) for each player in the NFL. To do this, we take
the following approach:

\begin{itemize}
\tightlist
\item
  estimate the value of each play (Section \ref{sec:ep_wp_model}),
\item
  estimate the effect of each player on play value added (Section
  \ref{sec:war_model}),
\item
  evaluate relative to replacement level (Section \ref{sec:repl_level}),
\item
  convert to a wins scale (Section \ref{sec:win_conversion}), and
\item
  and estimate the uncertainty in \textit{WAR} (Section \ref{sec:resample}).
\end{itemize}

This framework can be applied to any individual season, and we present
results for the 2017 season in Section \ref{results}. Due to data
restrictions, we currently are only able to produce \textit{WAR} estimates for
offensive skill position players. However, a benefit
of our framework is the ability to separate a player's total value into
the three components of \(WAR_{air}\), \(WAR_{yac}\), and
\(WAR_{rush}\). Additionally, we provide the first statistical estimates for a team's rush blocking based on play-by-play data.

\subsection{Division of Credit}
\label{sec:war_model}

In order to properly evaluate players, we need to allocate the portion
of a play's value \(\delta_{f,i}\) to each player on the field.
Unfortunately, the NFL does not publicly specify which players are on
the field for every play, preventing us from directly applying
approaches similar to those used in basketball and hockey discussed in
Section \ref{sec:prev-work-players}, where the presence of each player on
the playing surface is treated as an indicator covariate in a linear
model that estimates the marginal effect of that player on some game
outcome \citep{Kubatko07, Macdonald11, Thomas13}. Instead, the data
available publicly from the NFL and obtained via \texttt{nflscrapR} is
limited to only those players directly involved in the play, plus
contextual information about the play itself. For rushing plays, this
includes:

\begin{itemize}
\tightlist
\item
  Players: rusher and tackler(s)
\item
  Context: run gap (end, tackle, guard, middle) and direction (left,
  middle, right)
\end{itemize}

\begin{figure}[!h]
\includegraphics[width=14cm]{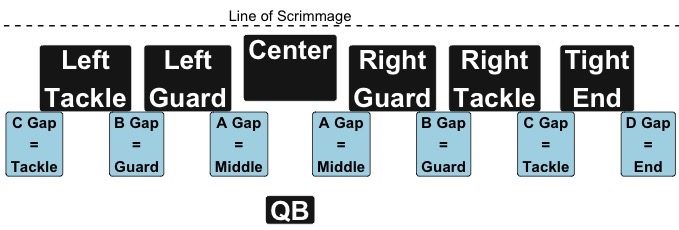}
\centering
\caption{Offensive Line Gaps for Rushing Plays.}
\label{fig:gaps}
\end{figure}

\autoref{fig:gaps} provides a diagram of the run gaps (in blue) and the
positions along the offensive line (in black). In the NFL play-by-play, the gaps are not referred to with letters, as they commonly are by football players and coaches; instead, the terms ``middle'', ``guard'', ``tackle'', and ``end'' are used. For the purposes of this paper, we define the following linkage between these two nomenclatures:

\begin{itemize}
\tightlist
\item
  ``A'' Gap = ``middle''
\item
  ``B'' Gap = ``guard''
\item
  ``C'' Gap = ``tackle''
\item
  ``D'' Gap = ``end''
\end{itemize}

For passing plays, information about each play includes:

\begin{itemize}
\tightlist
\item
  Players: passer, targeted receiver, tackler(s), and interceptor
\item
  Context: air yards, yards after catch, location (left, middle, right),
  and if the passer was hit on the play.
\end{itemize}

\hypertarget{multilevel-modeling}{%
\subsubsection{Multilevel Modeling}\label{multilevel-modeling}}

All players in the NFL belong to positional groups that dictate how they
are used in the context of the game. For example, for passing plays we
have the QB and the targeted receiver. However, over the course of an
NFL season, the average QB will have more pass attempts than the average
receiver will have targets, because there are far fewer QBs (more than
60 with pass attempts in the 2017 NFL season) compared to receivers
(more than 400 targeted receivers in the 2017 season).

Because of these systematic differences across positions, there are
differing levels of variation in each position's performance. Additionally, since every play involving the same player is
a repeated measure of performance, the plays themselves are not
independent.

To account for these structural features of football, we use a
multilevel model (also referred to as hierarchical, random-effects, or
mixed-effects model), which embraces this positional group structure and
accounts for the observation dependence. Multilevel models have recently
gained popularity in baseball statistics due to the development of
catcher and pitcher metrics by Baseball Prospectus
\citep{Brooks15, Turkenkopf15}, but have been used in sports dating back
at least to 2013 \citep{Thomas13}. Here, we novelly extend their use for
assessing offensive player contributions in football, using the play
values \(\delta_{f,i}\) from Section \ref{sec:ep_wp_model} as the
response.

In order to arrive at individual player effects we use
varying-intercepts for the groups involved in a play. A simple example
of modeling \(\delta_{f,i}\) with varying-intercepts for two groups, QBs
as \(Q\) and receivers as \(C\), with covariates $X_i$ and coefficients $\beta$ is as follows:
\begin{equation}
\label{ex_model_top_level}
\delta_{f,i} \sim Normal(Q_{q[i]} + C_{c[i]} + X_i \cdot \beta,\ \sigma_{\delta}^2),\ for\ i\ =\ 1,\hdots,n\  \mbox{plays},
\end{equation}

\noindent where the key feature distinguishing multilevel regression from
classical regression is that the group coefficients vary according to
their own model:
\begin{gather}
          Q_q \sim Normal(\mu_{Q},\ \sigma_{Q}^2),\ \mbox{for}\ q\ =\ 1,\hdots, \mbox{\#\ of\ QBs},\nonumber \\
          C_c \sim Normal(\mu_{C},\ \sigma_{C}^2),\ \mbox{for}\ c\ =\ 1,\hdots, \mbox{\#\ of\ receivers}.
\end{gather}

By assigning a probability distribution (such as the Normal distribution) to the group intercepts, \(Q_q\) and \(C_c\), with parameters estimated from the data (such as \(\mu_{Q}\) and \(\sigma_{Q}\) for passers), each estimate is pulled toward their respective group mean levels \(\mu_{Q}\) and \(\mu_{C}\). In this example, QBs and receivers involved in fewer plays will be pulled closer to their overall group averages as compared to those involved in more plays and thus carrying more information, resulting in partially pooled estimates \citep{Gelman07}. This approach provides us with average
individual effects on play value added while also providing the
necessary shrinkage towards the group averages. All models we use for
division of credit are of this varying-intercept form, and are fit using penalized likelihood via the \texttt{lme4} package in \texttt{R} \citep{lme4}. While these models are not explicitly Bayesian, as \citet{Gelman07} write, ``[a]ll multilevel models are Bayesian in the sense of assigning probability distributions to the varying regression coefficients", meaning we're taking into consideration all members of the group when estimating the varying intercepts rather than just an individual effect. 

Our assumption of normality for \(\delta_{f,i}\) follows from our focus on \(EPA\) and \(WPA\) values, which can be both positive and negative, exhibiting roughly symmetric distributions. We refer to an intercept estimating a player's average effect as their \emph{individual
points/probability added} (\(iPA\)), with points for modeling \(EPA\)
and probability for modeling \(WPA\). Similarly, an intercept estimating
a team's average effect is their \emph{team points/probability added}
(\(tPA\)). Tables \ref{table-pv-vars} and \ref{table-groups} provide the
notation and descriptions for the variables and group terms in the
models apportioning credit to players and teams on plays. The variables
in Table \ref{table-pv-vars} would be represented by \(X\), and their
effects by \(\beta\) in \autoref{ex_model_top_level}.

\begin{table}
\centering
\caption{Description of variables in the models assessing player and team effects.}
\label{table-pv-vars}
\begin{tabular}{ p{3cm} p{9cm}}
\hline \\ [-1.5ex]
Variable name & Variable description \\ [1ex]
\hline \\ [-1.5ex]
Home & Indicator for if the possession team was home \\ [1ex]
Shotgun & Indicator for if the play was in shotgun formation \\ [1ex]
NoHuddle & Indicator for if the play was in no huddle \\ [1ex]
QBHit & Indicator for if the QB was hit on a pass attempt \\  [1ex]
PassLocation & Set of indicators for if the pass location was either middle or right (reference group is left) \\ [1ex]
AirYards & Orthogonal distance in yards from the line of scrimmage to where the receiver was targeted or caught the ball \\ [1ex]
RecPosition & Set of indicator variables for if the receiver's position was either TE, FB, or RB (reference group is WR) \\ [1ex]
RushPosition & Set of indicator variables for if the rusher's position was either FB, WR, or TE (reference group is RB) \\ [1ex]
PassStrength & EPA per pass attempt over the course of the season for the possession team \\ [1ex]
RushStrength & EPA per rush attempt over the course of the season for the possession team \\ [1ex]
\hline
\end{tabular}
\end{table}

\begin{table}
\centering
\caption{Description of groups in the models assessing player and team effects.}
\label{table-groups}
\begin{tabular}{ p{2cm} p{2cm} p{8cm} }
\hline \\ [-1.5ex]
Group & Individual & Description \\ [1ex]
\hline \\ [-1.5ex]
$Q$ & $q$ & QB attempting a pass or rush/scramble/sack \\ [1ex]
$C$ & $c$ & Targeted receiver on a pass attempt \\ [1ex] 
$H$ & $\iota$ & Rusher on a rush attempt \\ [1ex]
$T$ & $\tau$ & Team-side-gap on a rush attempt, combination of the possession team, rush gap and direction \\ [1ex] 
$F$ & $\nu$ & Opposing defense of the pass \\ [1ex]
\hline
\end{tabular}
\end{table}

\subsubsection{Passing Models}

Rather than modeling the \(\delta_{f,i}\) (\(EPA\) or \(WPA\))
for a passing play, we take advantage of the availability of air yards
and develop two separate models for \(\delta_{f,i,air}\) and
\(\delta_{f,i,yac}\). We are not crediting the QB solely for the value
gained through the air, nor the receiver solely for the value gained
from after the catch. Instead, we propose that both the QB and receiver,
as well as the opposing defense, should have credit divided amongst them
for both types of passing values. We let \(\Delta_{air}\) and
\(\Delta_{yac}\) be the response variables for the air yards and yards
after catch models, respectively. Both models consider all passing
attempts, but the response variable depends on the model: 
\begin{gather}
\label{pass-value}
\Delta_{air} = \delta_{f,i,air} \cdot \boldsymbol{1}(\mbox{completion}) + \delta_{f,i} \cdot \boldsymbol{1}(\mbox{incompletion}), \nonumber \\
\Delta_{yac} = \delta_{f,i,yac} \cdot \boldsymbol{1}(\mbox{completion}) + \delta_{f,i} \cdot \boldsymbol{1}(\mbox{incompletion}),
\end{gather}

\noindent where \(\boldsymbol{1}(\mbox{completion})\) and
\(\boldsymbol{1}(\mbox{incompletion})\) are indicator functions for
whether or not the pass was completed. This serves to assign all
completions the \(\delta_{f,i,air}\) and \(\delta_{f,i,yac}\) as the
response for their respective models, while incomplete passes are
assigned the observed \(\delta_{f,i}\) for both models. In using this
approach, we emphasize the importance of completions, crediting accurate
passers for allowing their receiver to gain value after the catch.

The passing model for \(\Delta_{air}\) is as follows: 
\begin{gather}
          \Delta_{air} \sim Normal(Q_{air,q[i]} + C_{air,c[i]} + F_{air,\nu[i]} + \boldsymbol{A}_i \cdot \boldsymbol{\alpha},\ \sigma_{\Delta_{air}})\ \mbox{for}\ i\ =\ 1,\hdots,\ n\ \mbox{plays},\nonumber \\
          Q_{air,q} \sim Normal(\mu_{Q_{air}},\sigma_{Q_{air}}^2),\ \mbox{for}\ q\ =\ 1,\hdots, \mbox{\#\ of\ QBs},\nonumber \\
          C_{air,c} \sim Normal(\mu_{C_{air}},\sigma_{C_{air}}^2),\ \mbox{for}\ c\ =\ 1,\hdots, \mbox{\#\ of\ receivers}, \\
          F_{air,\nu} \sim Normal(\mu_{F_{air}},\sigma_{F_{air}}^2),\ \mbox{for}\ \nu\ =\ 1,\hdots, \mbox{\#\ of\ defenses},\nonumber
\end{gather}

\noindent where the covariate vector \(\boldsymbol{A}_i\) contains a set
of indicator variables for Home, Shotgun, NoHuddle, QBHit, Location,
RecPosition, as well as the RushStrength value while
\(\boldsymbol{\alpha}\) is the corresponding coefficient vector. The
passing model for \(\Delta_{yac}\) is of similar form: 

\begin{gather}
\Delta_{yac} \sim Normal(Q_{yac,q[i]} + C_{yac,c[i]} + F_{yac,\nu[i]} + \boldsymbol{B}_i \cdot \boldsymbol{\beta},\ \sigma_{\Delta_{yac}})\ \mbox{for}\ i\ =\ 1,\hdots,\ n\  \mbox{plays},\nonumber \\
Q_{yac,q} \sim Normal(\mu_{Q_{yac}},\ \sigma_{Q_{yac}}^2),\ \mbox{for}\ q\ =\ 1,\hdots, \mbox{\#\ of\ QBs},\nonumber \\
          C_{yac,c} \sim Normal(\mu_{C_{yac}},\ \sigma_{C_{yac}}^2),\ \mbox{for}\ c\ =\ 1,\hdots, \mbox{\#\ of\ receivers}, \\
          F_{yac,\nu} \sim Normal(\mu_{F_{yac}},\ \sigma_{F_{yac}}^2),\ \mbox{for}\ \nu\ =\ 1,\hdots, \mbox{\#\ of\ defenses},\nonumber
\end{gather}

\noindent where the covariate vector \(\boldsymbol{B}_i\) contains the
same set of indicator variables in \(\boldsymbol{A}_i\) but also
includes the AirYards and interaction terms between AirYards and the
various RecPosition indicators, with \(\boldsymbol{\beta}\) as its
respective coefficient vector. We include the RushStrength in the
passing models as a group-level predictor to control for the possession
team's rushing strength and the possible relationship between the two
types of offense. For QBs, their estimated \(Q_{air,q}\) and
\(Q_{yac,q}\) intercepts represent their \(iPA_{air}\) and \(iPA_{yac}\)
values respectively (same logic applies to receivers). Likewise, the
opposing defense values of \(F_{air,\nu}\) and \(F_{yac,\nu}\) are their
\(tPA_{air}\) and \(tPA_{yac}\) values.

\hypertarget{rushing-models}{%
\subsubsection{Rushing Models}\label{rushing-models}}

For rushing plays, we again model the play values \(\delta_{f,i}\).
However, we build two separate models, with one rushing model for QBs
and another for all non-QB rushes. This is because we cannot consistently separate
(in the publicly available data) designed QB rushes from scrambles on
broken plays, the characteristics of which result in substantially
different distributions of play value added. It is safe to assume all
non-QB rushes are designed rushes. Our rushing model for QBs consists of
all scrambles, designed runs, and sacks (to account for skilled rushing
QBs minimizing the loss on sacks). The QB rushing model is as follows:
\begin{gather}
          \delta_{f,i} \sim Normal(Q_{rush,q[i]} + F_{rush_{Q},\ \nu[i]} + \boldsymbol{\Gamma}_i \cdot \boldsymbol{\gamma},\ \sigma_{\delta_{rush_{Q}}})\ \mbox{for}\ i\ =\ 1,\hdots,\ n\ \mbox{plays},\nonumber \\
          Q_{rush,q} \sim Normal(\mu_{Q_{rush}},\ \sigma_{Q_{rush}}^2),\ \mbox{for}\ q\ =\ 1,\hdots, \mbox{\#\ of\ QBs}, \\
          F_{rush_{Q},\nu} \sim Normal(\mu_{F_{rush_{Q}}},\ \sigma_{F_{rush_{Q}}}^2),\ \mbox{for}\ \nu\ =\ 1,\hdots, \mbox{\#\ of\ defenses},\nonumber
\end{gather}

\noindent where the covariate vector \(\boldsymbol{\Gamma}_i\) contains
a set of indicator variables for Home, Shotgun, NoHuddle, as well as the
PassStrength variable where \(\boldsymbol{\gamma}\) is the corresponding
coefficient vector.

For the designed rushing plays of non-QBs, we include an additional
group variable \(T\). As detailed in \autoref{table-pv-vars} and \autoref{fig:gaps}, \(T\)
serves as a proxy for the offensive linemen or blockers involved in the
rushing attempt. Each team has seven possible \(T\) levels of the form
team-side-gap. For example, the Pittsburgh Steelers (PIT) have the
following levels: PIT-left-end, PIT-left-tackle, PIT-left-guard,
PIT-middle-center, PIT-right-guard, PIT-right-tackle, PIT-right-end. The
non-QB rushing model is as follows: 
\begin{gather}
          \delta_{f,i} \sim Normal(H_{\iota[i]} + T_{\tau[i]} + F_{rush,\nu[i]} + \boldsymbol{P}_i \cdot \boldsymbol{\rho},\ \sigma_{\delta_{rush}})\ \mbox{for}\ i\ =\ 1,\hdots,\ n\ \mbox{plays},\nonumber \\
          H_{\iota} \sim Normal(\mu_{H},\ \sigma_{H}^2),\ \mbox{for}\ \iota\ =\ 1,\hdots, \mbox{\#\ of\ rushers},\nonumber \\
          T_{\tau} \sim Normal(\mu_{T},\ \sigma_{T}^2),\ \mbox{for}\ \tau\ =\ 1,\hdots, \mbox{\#\ of\ team-side-gaps}, \\
          F_{rush,\nu} \sim Normal(\mu_{F_{rush}},\ \sigma_{F_{rush}}^2),\ \mbox{for}\ \nu\ =\ 1,\hdots, \mbox{\#\ of\ defenses},\nonumber
\end{gather}

\noindent where the covariate vector \(\boldsymbol{P}_i\) contains a set
of indicator variables for Home, Shotgun, NoHuddle, RushPosition, and
PassStrength, and where \(\boldsymbol{\rho}\) is the corresponding
coefficient vector. The resulting \(Q_{rush,q}\) and \(H_{rush,\iota}\)
estimates are the \(iPA_{rush}\) values for the QB and non-QB rushers,
respectively. Additionally, the \(T_{\tau}\) estimate is the
\(tPA_{rush,side-gap}\) for one of the seven possible side-gaps for the
possession team, while \(F_{rush,\nu}\) and \(F_{rush_{Q},\nu}\) are the
\(tPA_{rush}\) and \(tPA_{rush_Q}\) values for the opposing defense for
non-QB and QB rushes.

\subsubsection{Individual Points/Probability Added}
\label{sec:ipa}

Let \(\kappa\) refer to the number of attempts for a type of play. Using
an estimated type of \(iPA\) value for a player \(p\) and multiplying by
the player's associated number of attempts provides us with an
\emph{individual points/probability above average} (\(iPAA_p\)) value.
There are three different types of \(iPAA_p\) values for each position:
\begin{gather}
iPAA_{p,air} = \kappa_{p,pass} \cdot iPA_{p,air}, \nonumber \\
iPAA_{p,yac} = \kappa_{p,pass} \cdot iPA_{p,yac}, \\
iPAA_{p,rush} = \kappa_{p,rush} \cdot iPA_{p,air},\nonumber
\end{gather}

\noindent where the values for \(\kappa_{p,pass}\) and
\(\kappa_{p,rush}\) depend on the player's position. For QBs,
\(\kappa_{p,pass}\) equals their number of pass attempts, while
\(\kappa_{p,rush}\) is the sum of their rush attempts, scrambles, and
sacks. For non-QBs \(\kappa_{p,pass}\) equals their number of targets
and \(\kappa_{p,rush}\) is their number of rush attempts. Summing all
three components provides us with player \(p\)'s total individual
points/probability above average, \(iPAA_p\).

\hypertarget{comparing-to-replacement-level}{%
\subsection{Comparing to Replacement
Level}\label{comparing-to-replacement-level}}

\label{sec:repl_level}

As described in Section \ref{sec:prev-work-players}, it is desirable to
calculate a player's value relative to a ``replacement level'' player's
performance. There are many ways to define replacement level. For
example, \citet{Thomas15} define a concept called ``poor man's
replacement'', where players with limited playing time are pooled, and a
single effect is estimated in a linear model, which is considered
replacement level. Others provide more abstract definitions of
replacement level, as the skill level at which a player can be acquired
freely or cheaply on the open market \citep{Tango07}.

We take a similar approach to the \textit{openWAR} method, defining
replacement level by using a roster-based approach \citep{Baumer15}, and
estimating the replacement level effects in a manner similar to that of
\citet{Thomas15}. \citet{Baumer15} argue that ``replacement level''
should represent a readily available player that can replace someone
currently on a team's active roster. Due to differences in the number of
active players across positions in football, we define replacement level
separately for each position. Additionally, because of usage for the
different positions in the NFL, we find separate replacement level
players for receiving as compared to rushing. In doing so, we
appropriately handle cases where certain players have different roles.
For example, a RB that has a substantial number of targets but very few
rushing attempts can be considered a replacement level rushing RB, but
not a replacement level receiving RB.

Accounting for the 32 NFL teams and the typical construction of a roster
\citep{Lillibridge13}, we consider the following players to be ``NFL
level'' for each the non-QB positions:
\begin{itemize}
\tightlist
\item
  rushing RBs = \(32 \cdot 3 = 96\) RBs sorted by rushing attempts,
\item
  rushing WR/TEs = \(32 \cdot 1 = 32\) WR/TEs sorted by rushing attempts,
\item
  receiving RBs = \(32 \cdot 3 = 96\) RBs sorted by targets,
\item
  receiving WRs = \(32 \cdot 4 = 128\) WRs sorted by targets,
\item
  receiving TEs = \(32 \cdot 2 = 64\) TEs sorted by targets.
\end{itemize}

Using this definition, all players with fewer rushing attempts or
targets than the NFL level considered players are deemed
replacement-level. This approach is consistent with the one taken by
Football Outsiders \citep{Schatz03}. We combine the rushing replacement
level for WRs and TEs because there are very few WRs and TEs with
rushing attempts.

\begin{figure}[!h]
\includegraphics[width=14cm]{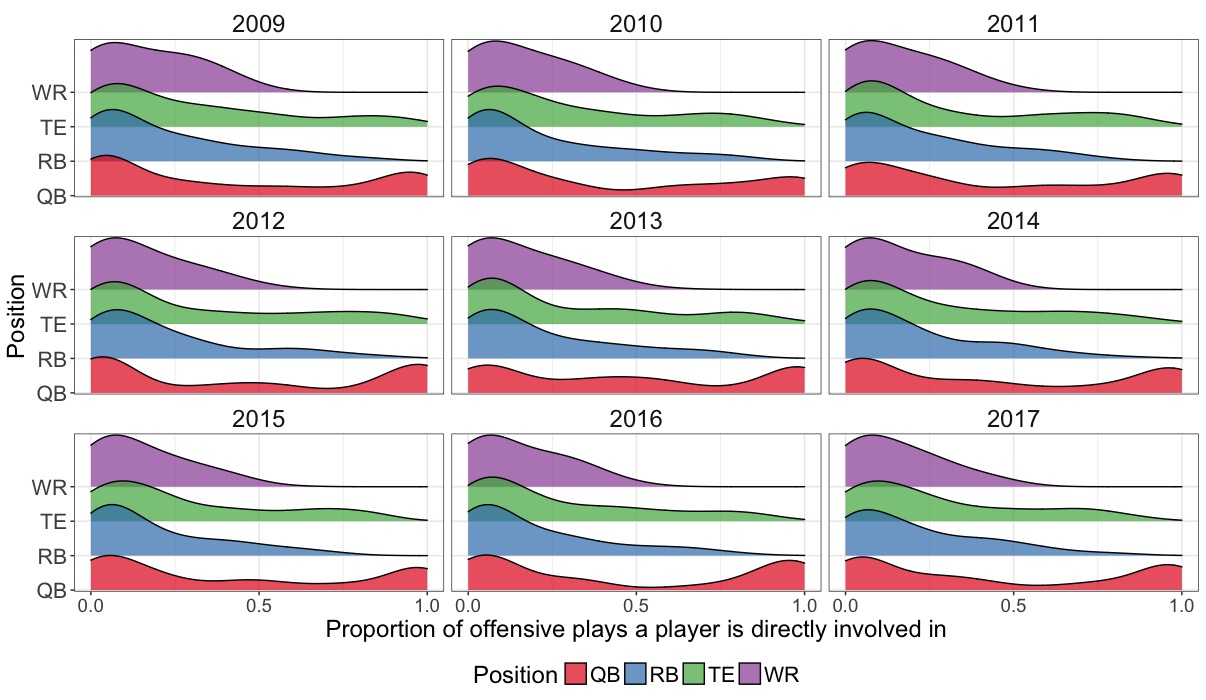}
\centering
\caption{Distribution of the proportion of offensive plays a player is directly involved in by position (2009-2017).}
\label{play-distr-pos}
\end{figure}

In order to find replacement level QBs, we proceed in a different
manner, due to the nature of QB usage in the NFL.
\autoref{play-distr-pos} displays the distribution of the percentage of
a team's plays in which a player is directly involved (passer, receiver,
or rusher) by position using data from 2009 to 2017. This does not represent the percentage of team snaps by a player, but
rather for a given position that is directly involved in a play, it shows
the distribution of team play percentages for every player of that
position (e.g. New Orleans Saints' RB Alvin Kamara was involved in
38.39\% of all Saints plays that directly involved a RB). While the
distributions for RB, WR, and TE are unimodal and clearly skewed right,
the distributions for QBs are bimodal for each season. This is an
unsurprising result, since most NFL teams rely on a single QB for an
entire season, resulting in them being involved in more than 80\% of the
team's plays at QB.

Observing this clear difference in the distribution for QBs, we consider
two definitions of replacement level for QBs. The first is to define a
replacement-level as any QB with less than ten percent involvement in
their team's plays that directly involve QBs. This approach essentially
asserts that backup QBs with limited playing time should represent
replacement level for QBs, rather than assuming all NFL teams have at
least a certain number of NFL level QBs on their roster. The second
option we consider is to limit NFL level to be the 32 QBs that attempted
a pass in the first quarter of the first game of the season for each
team, and label all remaining QBs as replacement level. The logic here
is that NFL teams typically do not sign free agent QBs outside of their
initial roster during the course of the season because it takes time to
learn a team's playbook and offensive schemes. We recognize that these
definitions are far from perfect, but we hope they provide a starting
point for defining replacement level from which researchers can improve
upon in the future.

Prior to fitting the models discussed in Section \ref{sec:war_model},
every player who is identified as replacement level is replaced in their
corresponding play-by-play data with their replacement label
(e.g.~Replacement QB, Replacement RB-rushing, Replacement RB-receiving,
etc). By doing so, all replacement level players for a particular
position and type (receiving versus rushing) have the same
\(iPA^{repl}\) estimate. We then calculate a player's value above
replacement, \emph{individual points/probability above replacement}
(\(iPAR_p\)) in the same manner as \citet{Baumer15} and
\citet{Thomas15}, by calculating a replacement level ``shadow'' for a
particular player. For a player \(p\), this is done by first calculating
their replacement ``shadow'' value, \(iPAA_p^{repl}\) by using their respective number of
attempts: 
\begin{gather}
  iPAA^{repl}_{p,air} = \kappa_{pass} \cdot iPA^{repl}_{air},\nonumber \\
  iPAA^{repl}_{p,yac} = \kappa_{pass} \cdot iPA^{repl}_{yac},\\
  iPAA^{repl}_{p,rush} = \kappa_{rush} \cdot iPA^{repl}_{rush},\nonumber
\end{gather}

\noindent which leads to natural calculations for the three \(iPAR\)
values: 
\begin{gather}
  iPAR_{p,air} = iPAA_{p,air} - iPAA^{repl}_{p,air},\nonumber \\
  iPAR_{p,yac} = iPAA_{p,yac} - iPAA^{repl}_{p,yac},\\
  iPAR_{p,rush} = iPAA_{p,rush} - iPAA^{repl}_{p,rush}.\nonumber
\end{gather}

\noindent Taking the sum of the three, we arrive at a player's total
\(iPAR_p\).

\subsection{Conversion to Wins}
\label{sec:win_conversion}

If the play's value used for modeling purposes was \(WPA\) based, then
the final \(iPAR\) values are an individual's win probability added above
replacement, which is equivalent to their \emph{wins above replacement}
(\(WAR\)).  However, for the \(EPA\)-based play value response, the \(iPAR\)
values represent the individual expected points added above replacement, and thus
require a conversion from points to wins.  We use a linear regression
approach, similar to that of \citet{Zhou17} for football and \citet{Thomas15} for hockey, to estimate the relationship
between a team \(t\)'s regular season win total and their score
differential (\(S\)) during the season, \begin{equation}
Wins_t = \beta_0 + \beta_{S}S_t + \epsilon_t,\text{ where } \epsilon_t \sim N(0, \sigma^2)\ (iid) .
\end{equation}

\begin{figure}[!h]
\includegraphics[width=14cm]{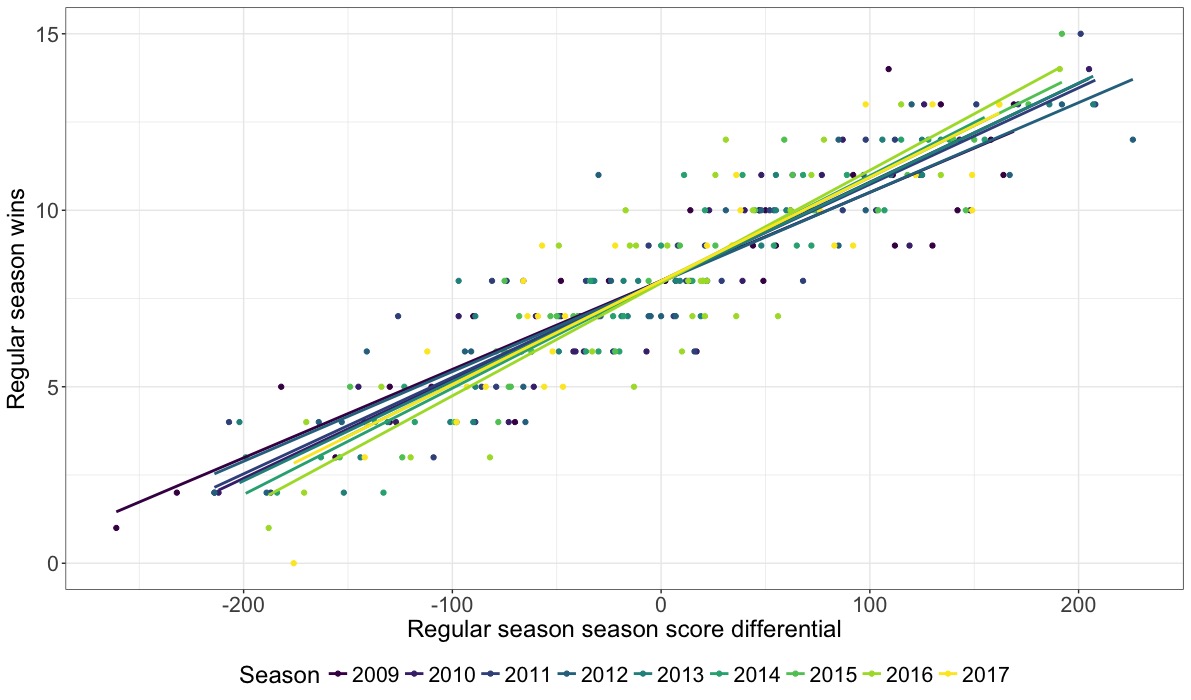}
\centering
\caption{Relationship between number of wins and score differential in the regular season by year (2009-2017).}
\label{wins-score}
\end{figure}

\autoref{wins-score} displays the estimated linear regression fits for
each season from 2009 to 2017. The resulting coefficient estimate
\(\hat{\beta}_{S}\) represents the increase in the number of
wins for each one point increase in score differential. Thus we take the
reciprocal, \(\frac{1}{\hat{\beta}_{S}}\) to arrive at the number of
points per win. We estimate \(WAR\) for the \(EPA\) based approach by
taking the \(iPAR\) values and dividing by the estimated points per win
(equivalent to multiplying \(iPAR\) by \(\hat{\beta}_{S}\)). 

\subsection{Uncertainty}
\label{sec:resample}

Similar to the approach taken by \citet{Baumer15} for estimating the
variability in their \emph{openWAR} metric, we use a resampling strategy
to generate distributions for each individual player's \(WAR\) values.
Rather than resampling plays in which a particular player is involved
to arrive at estimates for their performance variability, we resample
entire team drives.  We do this to account for the fact that player usage
is dependent on team decision making, meaning that the random variation
in individual events is dependent upon the random variation in team
events.  Thus, we must resample at the team level to account for the
variability in a player's involvement. The decision to resample whole
drives instead of plays is to represent sampling that is more realistic
of game flows due to the possibility of dependencies within a drive with
regards to team play-calling.  We recognize this is a simple viewpoint
of possible play correlations and consider exploration of this concept
for future work. In Section \ref{results}, all uncertainty estimation uses this drive-resampling approach, with 1000 simulated seasons.

\section{Results}
\label{results}

Given the definitions in Section \ref{sec:repl_level}, we found the following replacement level designations for the 2017 NFL season for non-QB positions:

\begin{itemize}
\tightlist
\item
  rushing:  52 of the 148 RBs are replacement level,
\item
  rushing:  278 of the 310 of the WR/TEs are replacement level,
\item
  receiving:  52 of the 148 RBs are replacement level
\item
  receiving:  73 of the 201 WRs are replacement level,
\item
  receiving:  45 of the 109 TEs are replacement level.
\end{itemize}

For the QB position, we consider both approaches discussed in
Section \ref{sec:repl_level}.  With the ``ten percent of QB plays cutoff'' approach resulting
in 25 replacement level QBs, and the ``one QB for each team'' approach resulting in 39 replacement level QBs out of the 71 in total.

\begin{figure}[!h]
\includegraphics[width=16cm]{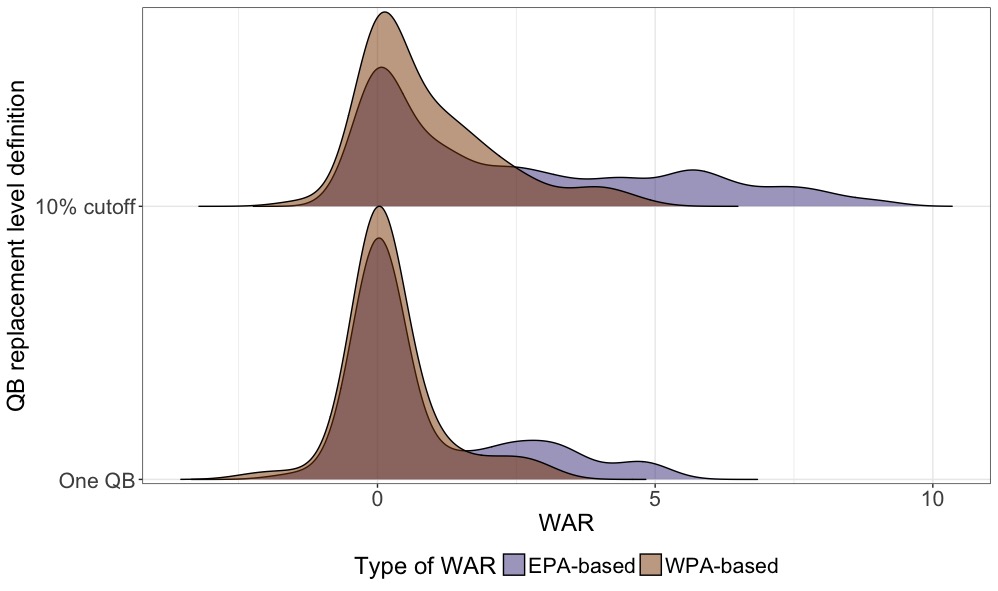}
\centering
\caption{Distribution of QB \textit{WAR} in 2017 season by type and replacement level definition.}
\label{qb-war-distr}
\end{figure}

First we compare the distributions of both types of \(WAR\) estimates, \(EPA\)-based and \(WPA\)-based, for the two considered definitions of replacement level QBs in \autoref{qb-war-distr}. It is clear that the ``one QB for each team'' approach for defining replacement level leads to lower \(WAR\) values in general, likely because some QBs who begin the season as back-ups perform better than those who begin the season as starters, yet are designated replacement level with this approach. For simplicity we only consider the ten percent cutoff rule for the rest of the paper.

\begin{figure}[!h]
\includegraphics[width=16cm]{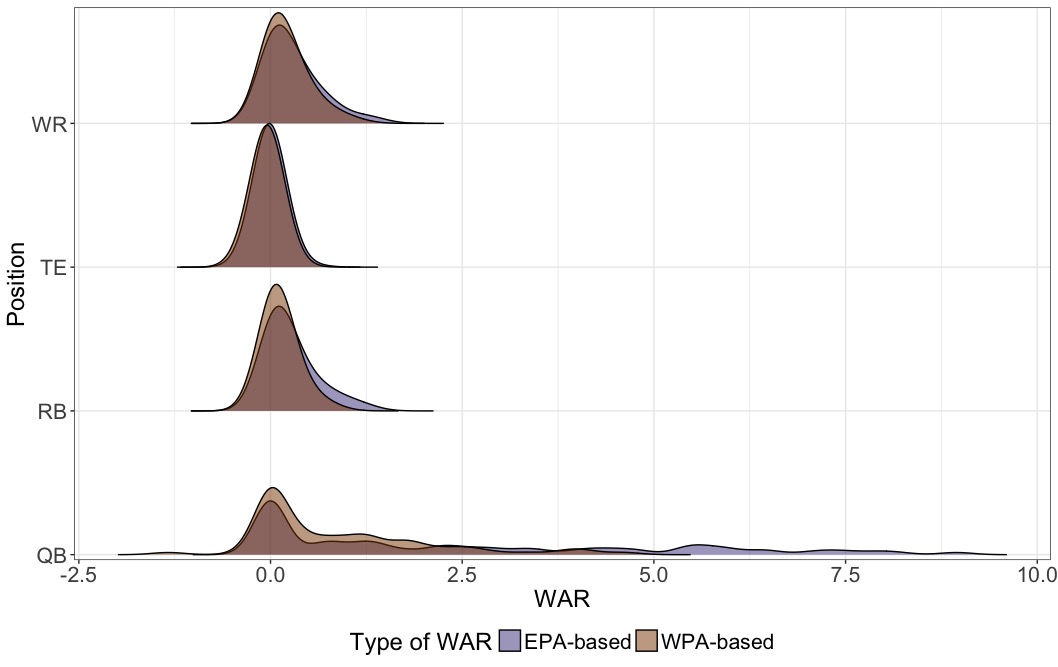}
\centering
\caption{Distribution of \textit{WAR} in 2017 season by type and position (ten percent cutoff used for replacement level QBs).}
\label{war-distr}
\end{figure}

We compare the distributions for both types of \(WAR\) estimates,
\(EPA\)-based and \(WPA\)-based, by position in \autoref{war-distr}. For
all positions, the \(EPA\)-based \(WAR\) values tend be higher than the
\(WPA\)-based values. This could be indicative of a player performing
well in meaningless situations due to the score differential,
particularly for QBs.  It is clear that QBs have larger \(WAR\) values
than the other positions, reflecting their involvement in every passing
play and potentially providing value by rushing.  Although this coincides with conventional wisdom regarding the importance of the QB position,
we note that we have not controlled for all possible contributing factors, such as the specific offensive linemen, the team's offensive schemes, or 
the team's coaching ability due to data limitations.  Researchers with
access to this information could easily incorporate their proprietary data into this framework to reach a better assessment of QB value. 

\begin{figure}[!h]
\includegraphics[width = 16cm]{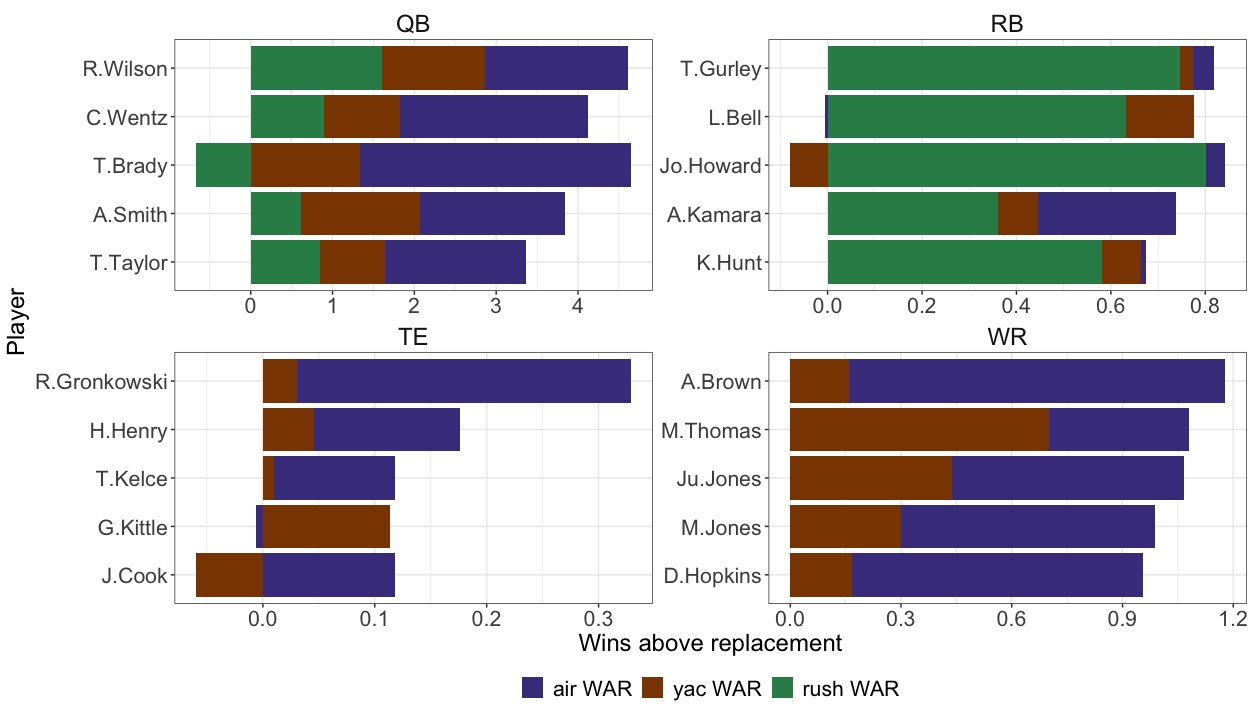}
\centering
\caption{Top five players in \textit{WAR} by position for the 2017 season.}
\label{war-bars}
\end{figure}

Following Major League Baseball's 2017 MVP race, \(WAR\) has received
heavy criticism for its unclear relationship with wins
\citep{James17, Tango17}.  For this reason, we focus on the \(WPA\)-based
version of \(WAR\), with its direct relationship to winning games.
\autoref{war-bars} displays the top five players based on total \(WAR\)
for each position in the 2017 season.  Each chart is arranged in
descending order by each player's estimated \textit{WAR}, and displays the three separate \(WAR\) values of \(WAR_{air}\), \(WAR_{yac}\) and
\(WAR_{rush}\). By doing this separation, we can see how certain types
of players vary in their performances. Tom Brady for instance is the only
QB in the top five with negative \(WAR_{rush}\).  Alvin Kamara
appears to be providing roughly equal value from rushing and receiving,
while the other top RB performances are primarily driven by rushing success.

\begin{figure}[!h]
\includegraphics[width=16cm]{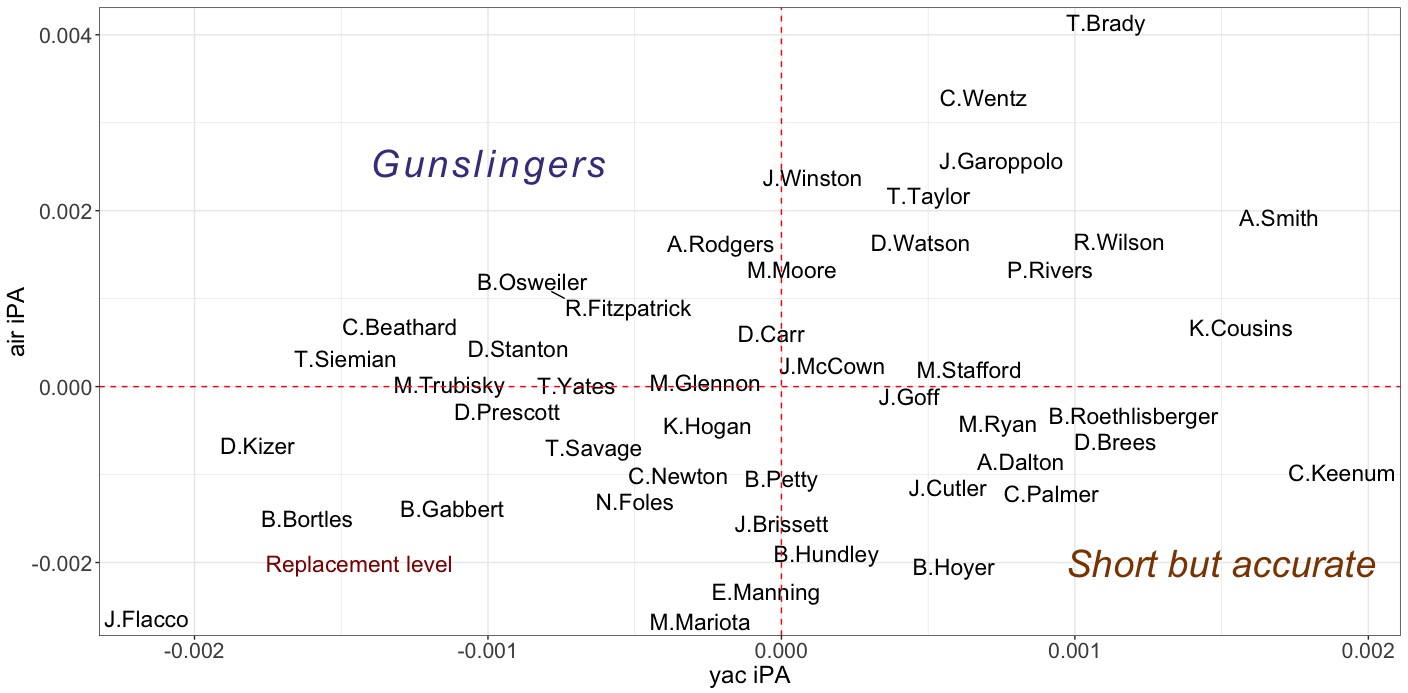}
\centering
\caption{Estimates for QB efficiency from \(iPA_{air}\) against \(iPA_{yac}\) for the 2017 season.}
\label{qb-ipa-plots}
\end{figure}

\begin{figure}[!h]
\includegraphics[width=16cm]{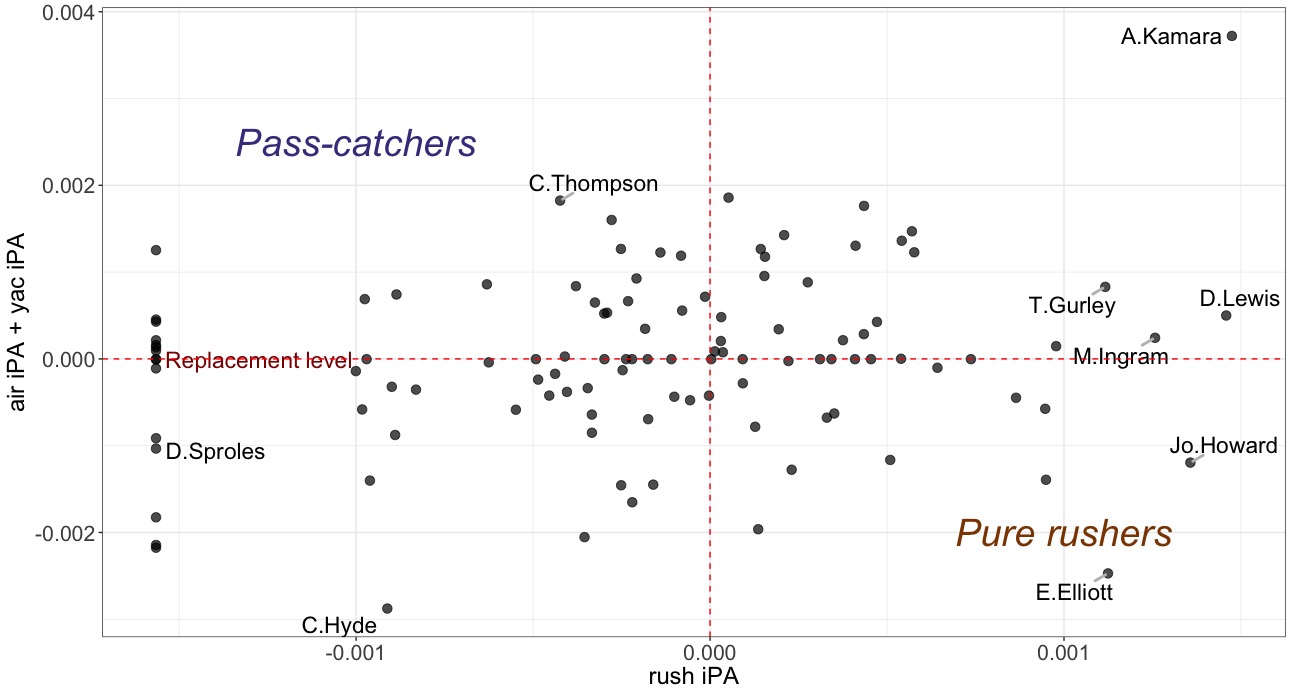}
\centering
\caption{Estimates for RB efficiency from receiving (\(iPA_{air}\ +\ iPA_{yac}\)) against rushing (\(iPA_{rush}\)) for the 2017 season.}
\label{rb-ipa-plots}
\end{figure}

Elaborating on this separation of types of players, we can use the random
intercepts from the multilevel models, the \(iPA\) values, to see the underlying structure of players in terms of their efficiency.  Figures \ref{qb-ipa-plots} and \ref{rb-ipa-plots}
reveals the separation of types of QBs and RBs respectively.  The origin point for both charts represents league averages.  For QBs, we plot their estimates for \(iPA_{air}\) against \(iPA_{yac}\), providing an overview of the types of passers in the NFL. The two components represent different skills of being able to provide value by throwing deep passes through the air, such as Jameis Winston, as compared to short but accurate passers such as Case Keenum.  We can also see where the replacement level QB estimates place for context.  For RBs, we add together their \(iPA_{air}\) and \(iPA_{yac}\) estimates to summarize their individual receiving effect and plot this against their \(iPA_{rush}\) estimates.  This provides a separation between RBs that provide value as receivers versus those who provide positive value primarily from rushing, such as Ezekiel Elliott.  New Orleans Saints RB Alvin Kamara stands out from the rest of the league's RBs, providing elite value in both areas.

\begin{figure}[!h]
\includegraphics[width=14cm]{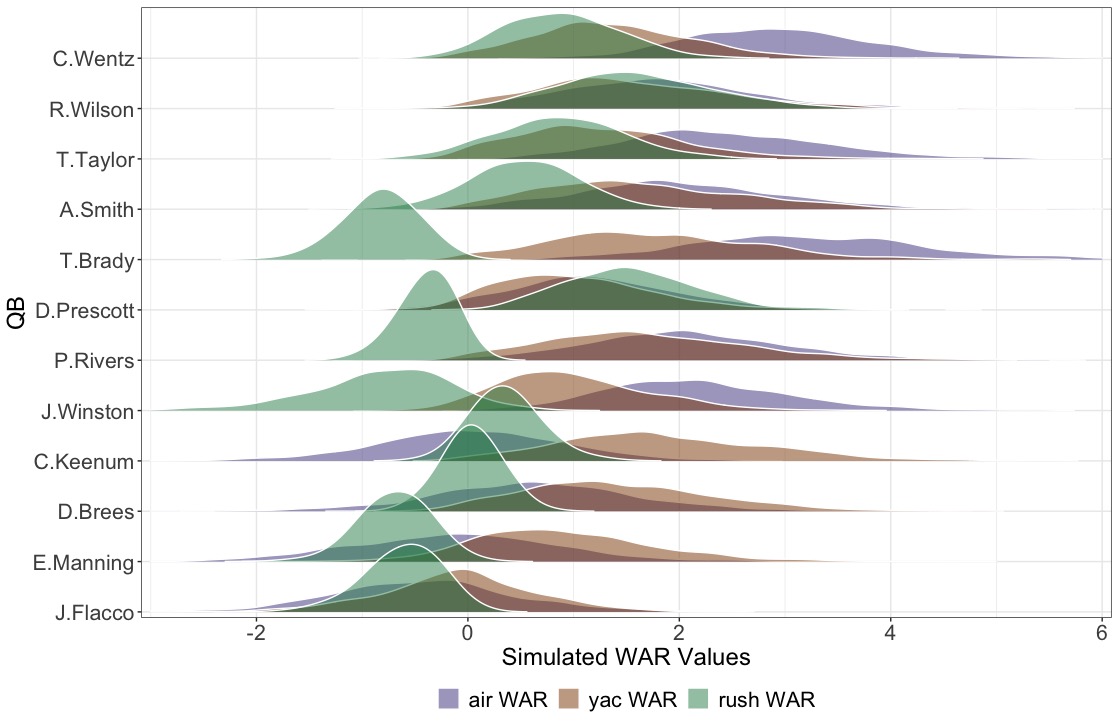}
\centering
\caption{Simulation distributions of 2017 \textit{WAR} by type for a selection of twelve QBs.}
\label{qb-war-sims}
\end{figure}

Using the drive resampling approach outlined in Section \ref{sec:resample},
we can compare the variability in player performance based on
1000 simulated seasons.  \autoref{qb-war-sims} compares the simulation
distributions of the three types of \(WAR\) values (\(WAR_{air}\),
\(WAR_{yac}\), \(WAR_{rush}\)) for selected QBs in the 2017 NFL season,
with a reference line at 0 for replacement-level.  We can clearly see
that the variability associated with player performance is not constant,
which is not suprising given the construction of the resampling at the
drive level.  However, we can see some interesting features of QB
performances, such as how Seattle Seahawks QB Russell Wilson's three
types of \(WAR\) distributions are overlapping significantly, emphasizing
his versatility.  Additionally, New England Patriots QB Tom Brady
displays large positive \(WAR_{air}\) and \(WAR_{yac}\) values, but a
clearly negative \(WAR_{rush}\) value.  Finally, Joe Flacco's 2017 performance was at or below replacement level in the vast majority of simulations across all three types of \textit{WAR}, indicating that he is not elite.

\begin{figure}[!h]
\includegraphics[width=14cm]{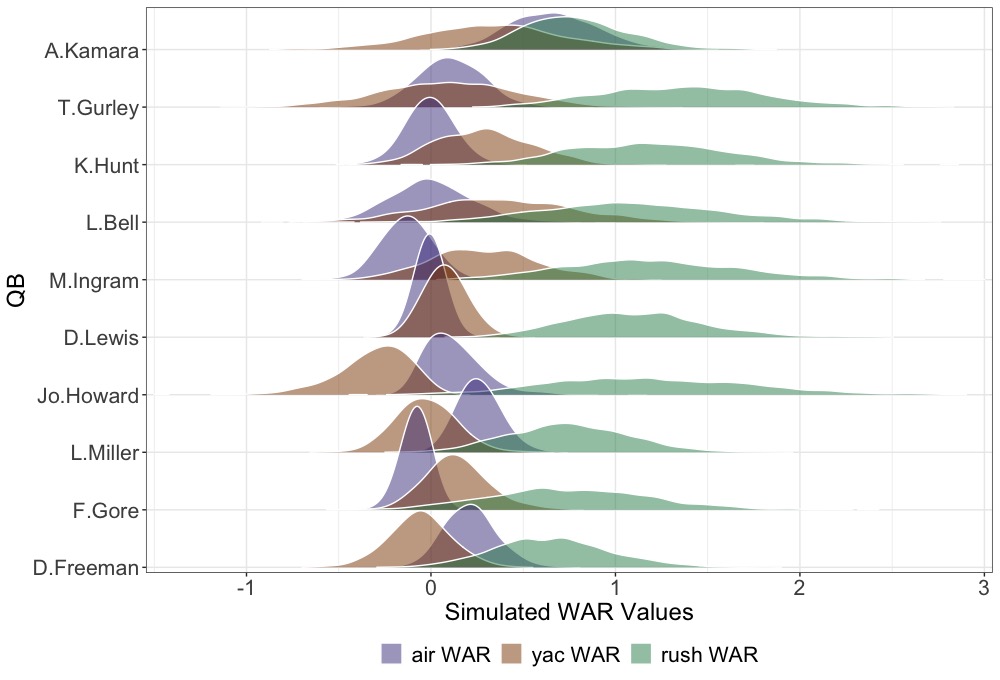}
\centering
\caption{Simulation distributions of 2017 \textit{WAR} value by type for top ten RBs.}
\label{rb-war-sims}
\end{figure}

\begin{figure}[!h]
\includegraphics[width=14cm]{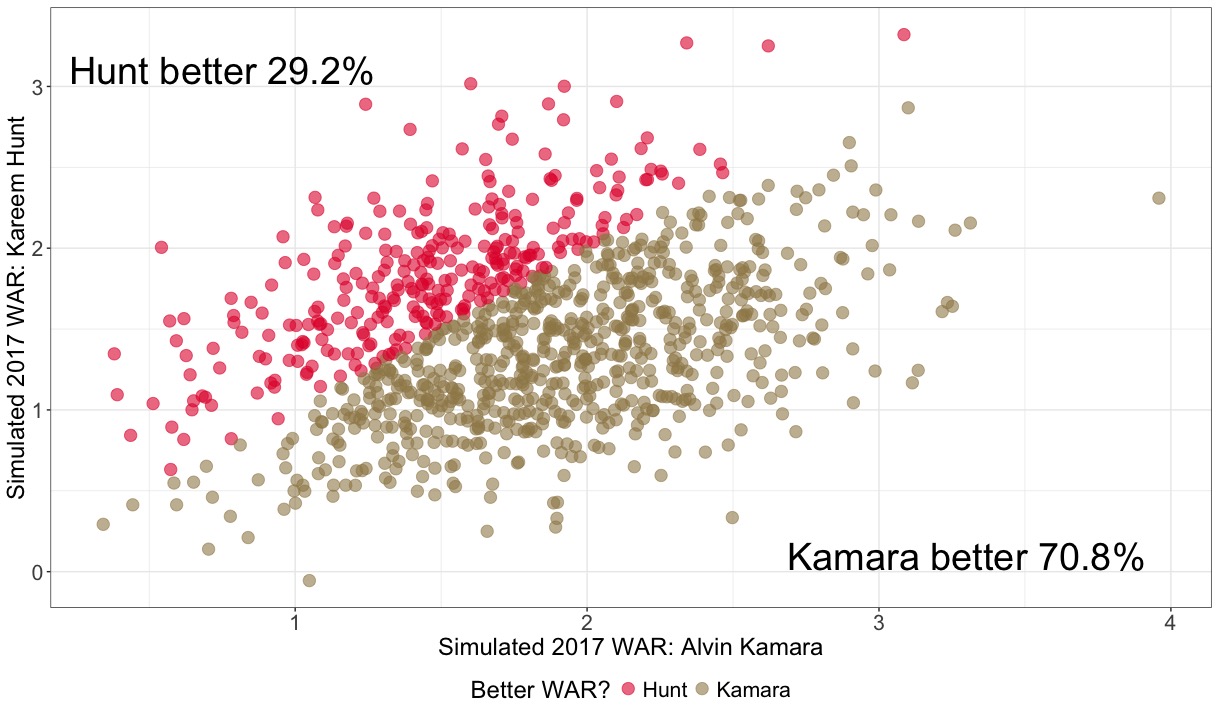}
\centering
\caption{Joint distribution of \textit{WAR} for Alvin Kamara vs. Kareem Hunt in 2017.}
\label{hunt-kam-comp}
\end{figure}

\autoref{rb-war-sims} displays the simulation distributions for the top
ten RBs during the 2017 NFL season, as ranked by their average total \(WAR\) across all simulations. Relative to the \(WAR\) values for QBs in \autoref{rb-war-sims}, the best RBs in the league are providing limited value to their teams. This is in agreement with the recent trend of NFL teams, who have been paying QBs increasing salaries but compensating RBs less \citep{Morris17}. Two of
the top RBs in the 2017 were rookies Alvin Kamara and Kareem Hunt,
resulting into discussion of which player deserved to be the NFL's
rookie of the year. Similar to \citet{Baumer15} we address this question
using our simulation approach and display the joint distribution of the
two player's 2017 performances in \autoref{hunt-kam-comp}. In nearly
71\% of the simulated seasons, Kamara leads Hunt in \(WAR\) providing us
with reasonable certainty in Kamara providing more value to his team
than Hunt in his rookie season. It should not come as a surprise that
there is correlation between the player performances as each simulation
consists of fitting the various multilevel models resulting in new
estimates for the group averages, individual player intercepts as well
as the replacement level performance.

Additionally, we examine the consistency of the \(WPA\)-based \(WAR\)
from season-to-season based on the autocorrelation within players
between their 2016 and 2017 seasons (excluding replacement level) and
compare this to other commonly used statistics for QBs and RBs. Seen in
\autoref{table-qb-stats}, our estimates for QB \(WAR\) displayed higher
correlations than both the commonly used Passer Rating statistic as well
as Pro-Football-Reference.com's Adjusted Net Yards per Passing Attempt
(ANY/A), which includes yards lost from sacks \citep{PFR}.  We also see
higher correlations for RB \(WAR\) as compared to Brian Burke's Success
Rate (percentage of rush attempts with \(EPA\) greater than zero) and
rushing yards per attempt.  Future work should consider a proper review
and assessment of football statistics accounting for the number of
attempts needed for determing the reliability of a statistic as well as
accounting for when a player changes teams \citep{Yurko17}, and also
apply the framework laid out by \citet{Franks17}.

\begin{table}
\centering
\caption{Autocorrelation of QB statistics between 2016-17 seasons.}
\label{table-qb-stats}
\begin{tabular}{ c c c c}
\hline \\ [-1.5ex]
 & \textit{WAR} & Passer Rating & ANY/A \\ [1ex]
\hline \\ [-1.5ex]
Autocorrelation & 0.598 & 0.478 & 0.295 \\ [1ex]
\hline
\end{tabular}
\end{table}

\begin{table}
\centering
\caption{Autocorrelation of RB statistics between 2016-17 seasons.}
\label{table-rb-stats}
\begin{tabular}{ c c c c}
\hline \\ [-1.5ex]
 & \textit{WAR} & Success Rate & Yards per Attempt \\ [1ex]
\hline \\ [-1.5ex]
Autocorrelation & 0.431 & 0.314 & 0.337 \\ [1ex]
\hline
\end{tabular}
\end{table}

\begin{figure}[!h]
\includegraphics[width=14cm]{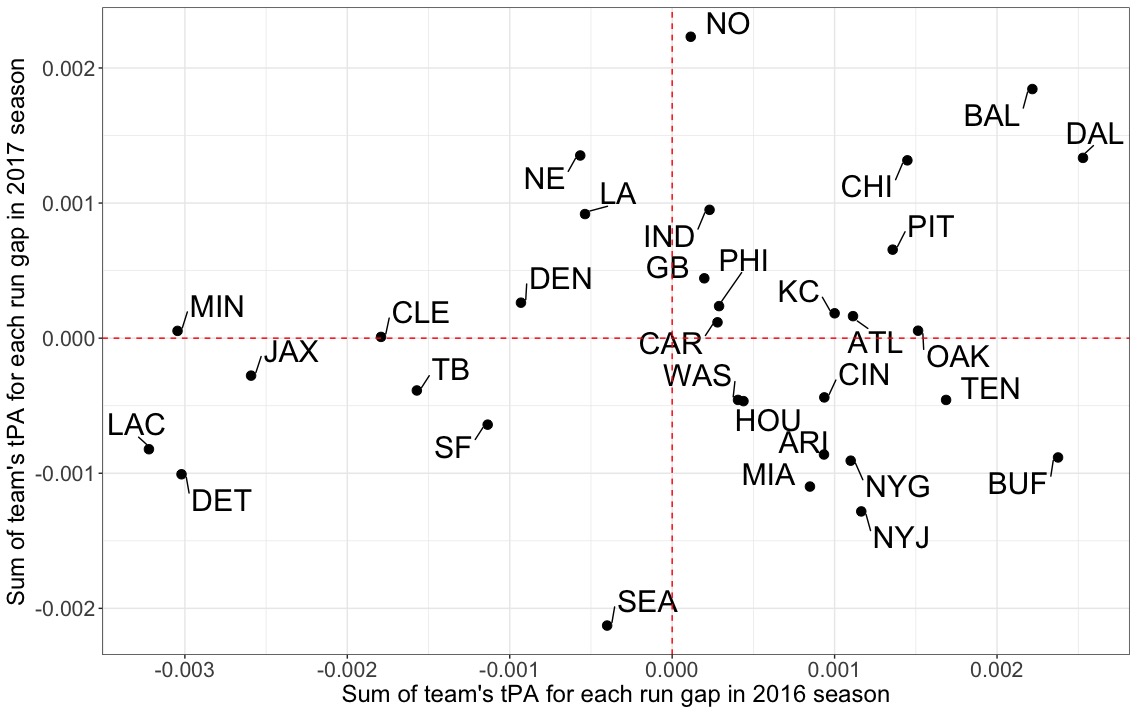}
\centering
\caption{Team offensive line measures.}
\label{oline-plot}
\end{figure}

Although it does not provide a measure for individual players'
contributions, we can sum together the seven possible
\(tPA_{rush,side-gap}\) estimates for a team providing a proxy for their
offensive line's overall efficiency in contributing to rushing plays.
We can also look at individual side-gaps for specific teams to assess
their offensive line's performance in particular areas.  
\autoref{oline-plot} displays the \(tPA_{rush,side-gap}\) sum in 2017
against 2016 for each NFL team.  The red lines provide indication to
average performances in each year, so teams in the upper right quadrant
performed above average overall in both years such as the Dallas Cowboys
(DAL) which are known to have one of the best offensive lines in
football.

\section{Discussion and Extensions}
\label{sec:discussion}

In this work, we have provided four major contributions to the statistical analysis of NFL football, in areas that can impact both on-field and player personnel decisions.  These contributions are broken into three categories:  software development and data, play evaluation, and player evaluation.

\subsection{Data and Software Development}

In the area of data access and software development, we provide an \texttt{R} package, \texttt{nflscrapR}, to provide easy access to publicly available NFL play-by-play data for researchers to use in their own analyses of the NFL.  This package has already been used by researchers to further research into NFL decision-making \citep{Yam18}.

\subsection{Novel Statistical Methods for Play Evaluation}

In the area of play evaluation, we make two contributions.  First, we introduce a novel approach for estimating expected points using a multinomial logistic regression model.  By using this classification approach, we more appropriately model the ``next score" response variable, improving upon previous approaches.  Additionally, our approach is fully reproducible, using only data provided publicly by the NFL.  Second, we use a generalized additive model for estimating in-game win probability, incorporating the results of the expected points model as input.  With these two play evaluation models, we can calculate measures such as expected points added and win probability added, which are commonly used to evaluate both plays and players.  

With the notable exception of \citet{Lock14}, researchers typically only vaguely discuss the methodology used for modeling expected points and/or win probability.  Additionally, prior researchers in this area typically do not provide their specific expected points and win probability estimates publicly for other researchers to use and explore. Recently, Pro Football Focus used our approach for modeling expected points and found a clear relationship between their player grades and expected points added \citep{Douglas17}.  Importantly, in our work, all of these measures are included directly into the play-by-play data provided by \texttt{nflscrapR}, and our methodology is fully described in this paper.  Moreover, all code used to build these expected points and win probability models is provided in \texttt{nflscrapR} and available on GitHub \texttt{https://github.com/ryurko/nflscrapR-models}.  By taking these important steps, we ensure that all of our methods are fully reproducible, and we make it as easy as possible for researchers to use, explore, and improve upon our work.

\subsection{Novel Statistical Methods for Player Evaluation}

In the area of player evaluation, we introduce several metrics for evaluating players via our \textit{nflWAR} framework.  We use multilevel models to isolate offensive skill player contribution and estimate their individual wins above replacement.  There are several key pieces of our \textit{WAR} estimation that merit discussion.  

First, estimates of \textit{WAR} are given for several different areas of the game, including passing through the air, passing for yards after the catch, rushing, receiving through the air, and receiving yards after the catch.  By compartmentalizing our estimates of player \textit{WAR}, we are able to better characterize players and how they achieved success.  For example, New Orleans Saints RB Alvin Kamara was unique in his success as both a rusher and a receiver in the 2017 NFL season, while other RBs like Los Angeles Rams RB Todd Gurley achieved most of their success as a rusher.  Similarly, Seattle Seahawks QB Russell Wilson was unique in his success as a rusher as well as from passing through the air and for yards after the catch, with about equal \textit{WAR} contributions in all three areas in the 2017 NFL season.  This is in contrast to New England QB Tom Brady, who had tremendous success passing through the air and passing for yards after the catch, but provided negative \textit{WAR} contributions as a rusher.  We are also able to characterize players like Case Keenum, who in the 2017 NFL season performed very well as a passer for yards after the catch, but not as well as a passer through the air. While these findings may not surprise knowledgeable football fans our framework also reveals the value of potentially overlooked skills such as the rushing ability of Tyrod Taylor and Dak Prescott, as seen in \autoref{qb-war-sims}. Their rushing value reflects not just their ability to scramble for positive value, but indicative of how they limit the damage done on sacks. Importantly, our player evaluation metrics are available for all skill position players, not just for QBs like previous approaches.

Second, our multilevel modeling approach allows us to estimate \textit{WAR} contributions for NFL offensive lines and their specific sides and gaps on rushing plays, providing the first statistical estimate of offensive line ability that also controls for factors such as RB ability, opposing defense, etc.  We recognize, however, that this is not a perfect measure of offensive line performance for a few reasons.  First, this does not necessarily capture individual linemen, as blocking can consist of players in motion and the involvement of other positions.  Second, there is likely some selection bias that is not accounted for in the play-by-play data that could influence specific side-gap estimates.  For example, a RB may cut back and find a hole on the left side of the line on a designed run to the right because there is nothing open on the right side, resulting in a play being scored as a run to the left.  Because of this selection bias at the RB level -- RBs are more often going to run towards holes and away from defenders -- our team-side-gap estimates may be biased, especially for teams with particularly strong or weak areas of their line.  This is an issue with the play-by-play data that likely cannot be remedied publicly until player-tracking data is made available by the NFL.  Finally, we lack information about which specific offensive linemen are on the field or even involved in plays, preventing us from fitting player-specific terms in our multilevel model that would provide \textit{WAR} estimates for individual offensive linemen.  Researchers with access to this data can build this into our modeling framework with minimal issues.  However, until more data becomes available, researchers can incorporate these measures with more nuanced approaches of measuring offensive line performance such as \citet{Alamar08} and \citet{Alamar11}.

Third, by adopting a resampling procedure similar to that of \citet{Baumer15}, we provide estimates of uncertainty on all \textit{WAR} estimates.  Our approach resamples at the drive-level, rather than resampling individual plays, to preserve the effects of any within-drive factors, such as play sequencing or play-calling tendencies.

Finally, our \textit{WAR} models are fully reproducible, with all data coming directly from \texttt{nflscrapR}, and with all code provided on GitHub \texttt{https://github.com/ryurko/nflWAR}.  Because we use parametric models, it is trivially easy to incorporate more information, such as information about which players are on the field, or information from player-tracking data.  We encourage future researchers to expand and improve upon our models in future work.

\subsection{The Road to WAR for Players at All Positions}
\label{sec:road_to_war}

One key benefit to our approach is that it can easily be augmented with the inclusion of additional data sources, e.g. player-tracking data or proprietary data collected by NFL teams.  One important way in which our models can be augmented comes via the inclusion of data about which players are present on the field for each play.  

Given this information, we can update our multilevel models from Section \ref{sec:war_model} by including additional positional groups.  For example, for the non-QB rushing model, we can update the model as follows:
\begin{gather}
          \delta_{f,i} \sim Normal(\sum_k O^{k}_{rush,\nu_k[i]} + \sum_g D^{g}_{rush,\nu_g[i]} + \boldsymbol{P}_i \cdot \boldsymbol{\rho},\ \sigma_{\delta_{rush}})\ \mbox{for}\ i\ =\ 1,\hdots,\ n\ \mbox{plays},\nonumber \\
          O^{k}_{rush,\nu_k} \sim Normal(\mu_{O^{k}_{rush}},\ \sigma_{O^{k}_{rush}}^2), \nonumber \\
          D^{g}_{rush,\nu_g} \sim Normal(\mu_{D^{g}_{rush}},\ \sigma_{D^{g}_{rush}}^2), \nonumber
\end{gather}

\noindent where \(O^{k}_{rush,\nu_k}\) are the intercepts for offensive positions (indexed by $k$ and varying according to their own model), \(D^{g}_{rush,\nu_g}\) are the intercepts for defensive positions (indexed by $g$ and varying according to their own model), and \(\boldsymbol{P}_i\) and \(\boldsymbol{\rho}\) are described as above.  Similar updates can be made to the models representing QB rushing, passing through the air, and passing for yards after catch.  After doing so, we can trivially calculate the individual points/probability above average for any player at any position following the approach outlined in Section \ref{sec:ipa}.  From there, we simply need adequate definitions for replacement level players at each of these positions, and we will have statistical \textit{WAR} estimates for players of any position, including all offensive players and all defensive players.

The data necessary for employing this approach \emph{does exist}, but it is not available publicly, and there are heavy restrictions on the uses of such data.  For example, Sportradar has a data product for the NFL called ``Participation Data'', which specifies all players present on the field for all plays, with data provided from the NFL.  This is stated directly:  ``Participation Data is complementary data collected by the NFL that indicates all 22 players on the field for every play of every game'' \citep{sportradar_data}.  

However, Sportradar's data sharing agreement explicitly prohibits the use of this data in the creation of new metrics, even if only used privately, as detailed in clauses 1.6 and 14.2 of the agreement \citep{sportradar_agreement}.  When colleagues reached out to Sportradar for clarification on potential data availability, they were told that there is no data sharing agreement for academic use, and that even if one were to purchase these data products, no new statistics or evaluation methods could be developed using this data, as per their terms and conditions.  It is not clear if the same restrictions would apply to NFL teams.

\subsection{Extensions Relevant to NFL Teams}

\texttt{nflscrapR} provides play-by-play data, including expected points and win probability results, dating back to 2009, and improvements are underway to extend this back even further.  As such, we can calculate player \textit{WAR} dating back to at least 2009.  If teams are able to implement the framework discussed in Section \ref{sec:road_to_war}, they would then have \textit{WAR} estimates for players at all positions dating back almost a full decade.  Teams that are able to do this could potentially gain substantial advantages in important areas of roster construction.

First, teams could more appropriately assess the contract values of players in free agency, similar to what is commonly done in baseball \citep{Paine15b}.

Second and perhaps most importantly, teams would be able to substantially improve their analysis for the NFL draft.  Using an approach similar to that of \citet{Citrone17}, teams could substitute an objective measure of \textit{WAR} in place of the more subjective measure of ``approximate value'' (AV) \citep{PFR}, in order to project the future career value in terms of \textit{WAR} for all players available in the NFL draft.  Additionally, teams employing this approach could create updated, \textit{WAR}-based versions of the ``draft pick value chart'', first attributed to Jimmy Johnson and later improved by \citet{Meers11} and \citet{Citrone17}.  In doing so, teams could more accurately assess the value of draft picks and potentially exploit their counterparts in trades involving draft picks.

\section*{Acknowledgements}

First and foremost, we thank the faculty, staff, and students in Carnegie Mellon University's Department of Statistics \& Data Science for their advice and support throughout this work.  We thank the now-defunct CMU Statistics NFL Research Group; the CMU Statistics in Sports Research Group; the Carnegie Mellon Sports Analytics Club; and the Carnegie Mellon Statistics Clustering, Classification, and Record Linkage Research Group for their feedback and support at all stages of this project.  In particular, we thank Devin Cortese, who provided the initial work in evaluating players with expected points added and win probability added, and Nick Citrone, whose feedback was invaluable to this project.  We thank Jonathan Judge for his insight on multilevel models.  We thank Michael Lopez and Konstantinos Pelechrinis for their help on matters relating to data acquisition and feedback throughout the process.  We thank Konstantinos Pelechrinis, the organizers of the Cascadia Symposium for Statistics in Sports, the organizers of the 6th Annual Conference of the Upstate New York Chapters of the American Statistical Association, the organizers of the Great Lakes Analytics in Sports Conference, the organizers of the New England Symposium on Statistics in Sports, and the organizers of the Carnegie Mellon Sports Analytics Conference for allowing us to present earlier versions of this work at their respective meetings; we thank the attendees of these conferences for their invaluable feedback.  We thank Jared Lander for his help with parts of \texttt{nflscrapR}.  Finally, we thank Rebecca Nugent for her unmatched dedication to statistical education, without which none of the authors would be capable of producing this work.

\bibliographystyle{DeGruyter}

\begin{thebibliography}{78}
\newcommand{\enquote}[1]{``#1''}
\providecommand{\natexlab}[1]{#1}
\providecommand{\url}[1]{\texttt{#1}}
\providecommand{\urlprefix}{URL }

\bibitem[{Alamar(2010)}]{Alamar10}
Alamar, B. (2010): \enquote{Measuring risk in nfl playcalling,} \emph{Journal
  of Quantitative Analysis in Sports}, 6.

\bibitem[{Alamar and Goldner(2011)}]{Alamar11}
Alamar, B. and K.~Goldner (2011): \enquote{The blindside project: Measuring the
  impact of individual offensive linemen,} \emph{CHANCE}, 24, 25--29.

\bibitem[{Alamar and Weinstein-Gould(2008)}]{Alamar08}
Alamar, B. and J.~Weinstein-Gould (2008): \enquote{Isolating the effect of
  individual linemen on the passing game in the national football league,}
  \emph{Journal of Quantitative Analysis in Sports}, 4.

\bibitem[{Albert(2006)}]{Albert06}
Albert, J. (2006): \enquote{Pitching statistics, talent and luck, and the best
  strikeout seasons of all-time,} \emph{Journal of Quantitative Analysis in
  Sports}, 2.

\bibitem[{Balreira et~al.(2014)Balreira, Miceli, and Tegtmeyer}]{Balreira14}
Balreira, E.~C., B.~K. Miceli, and T.~Tegtmeyer (2014): \enquote{An oracle
  method to predict nfl games,} \emph{Journal of Quantitative Analysis in
  Sports}, 10.

\bibitem[{Bates et~al.(2015)Bates, M{\"a}chler, Bolker, and Walker}]{lme4}
Bates, D., M.~M{\"a}chler, B.~Bolker, and S.~Walker (2015): \enquote{Fitting
  linear mixed-effects models using {lme4},} \emph{Journal of Statistical
  Software}, 67, 1--48.

\bibitem[{Baumer and Badian-Pessot(2017)}]{Baumer17}
Baumer, B. and P.~Badian-Pessot (2017): \enquote{Evaluation of batters and base
  runners,} in J.~Albert, M.~E. Glickman, T.~B. Swartz, and R.~H. Koning, eds.,
  \emph{Handbook of Statistical Methods and Analyses in Sports}, Boca Raton,
  Florida: CRC Press, 1--37.

\bibitem[{Baumer et~al.(2015)Baumer, Jensen, and Matthews}]{Baumer15}
Baumer, B., S.~Jensen, and G.~Matthews (2015): \enquote{openwar: An open source
  system for evaluating overall player performance in major league baseball,}
  \emph{Journal of Quantitative Analysis in Sports}, 11.

\bibitem[{Baumer and Matthews(2017)}]{Baumer17b}
Baumer, B. and G.~Matthews (2017): \emph{teamcolors: Color Palettes for Pro
  Sports Teams}, \urlprefix\url{http://github.com/beanumber/teamcolors}, r
  package version 0.0.1.9001.

\bibitem[{Becker and Sun(2016)}]{Becker16}
Becker, A. and X.~A. Sun (2016): \enquote{An analytical approach for fantasy
  football draft and lineup management,} \emph{Journal of Quantitative Analysis
  in Sports}, 12.

\bibitem[{Burke(2009)}]{Burke_EP}
Burke, B. (2009): \enquote{Expected point values,}
  \urlprefix\url{http://archive.advancedfootballanalytics.com/2009/12/expected-point-values.html}.

\bibitem[{Burke(2014)}]{BurkeEP}
Burke, B. (2014): \enquote{Expected points and expected points added
  explained,}
  \urlprefix\url{http://www.advancedfootballanalytics.com/index.php/home/stats/stats-explained/expected-points-and-epa-explained}.

\bibitem[{Carroll et~al.(1988)Carroll, Palmer, Thorn, and
  Pietrusza}]{Carroll88}
Carroll, B., P.~Palmer, J.~Thorn, and D.~Pietrusza (1988): \emph{The Hidden
  Game of Football}, New York, New York: Total Sports, Inc.

\bibitem[{Carter and Machol(1971)}]{Carter71}
Carter, V. and R.~Machol (1971): \enquote{Operations research on football,}
  \emph{Operations Research}, 19, 541--544.

\bibitem[{Causey(2013)}]{Causey13}
Causey, T. (2013): \enquote{Building a win probability model part 1,}
  \urlprefix\url{http://thespread.us/building-a-win-probability-model-part-1.html}.

\bibitem[{Causey(2015)}]{Causey15}
Causey, T. (2015): \enquote{Expected points part 1: Building a model and
  estimating uncertainty,}
  \urlprefix\url{http://thespread.us/expected-points.html}.

\bibitem[{Citrone and Ventura(2017)}]{Citrone17}
Citrone, N. and S.~L. Ventura (2017): \enquote{A statistical analysis of the
  nfl draft: Valuing draft picks and predicting future player success,}
  Presented at the Joint Statistical Meetings.

\bibitem[{Clark et~al.(2013)Clark, Johnson, and Stimpson}]{Clark13}
Clark, T.~K., A.~W. Johnson, and A.~J. Stimpson (2013): \enquote{Going for
  three: Predicting the likelihood of field goal success with logistic
  regression,} \emph{MIT Sloan Sports Analytics Conference}.

\bibitem[{Dasarathy(1991)}]{Dasarathy}
Dasarathy, B.~V. (1991): \emph{Nearest neighbor (NN) norms: NN pattern
  classification techniques}, Los Alamitos, CA: IEEE Computer Society Press.

\bibitem[{Deshpande and Jensen(2016)}]{Deshpande16}
Deshpande, S.~K. and S.~T. Jensen (2016): \enquote{Estimating an nba player's
  impact on his team's chances of winning,} \emph{Journal of Quantitative
  Analysis in Sports}, 12.

\bibitem[{Douglas and Eager(2017)}]{Douglas17}
Douglas, B. and E.~A. Eager (2017): \enquote{Examining expected points and
  their interactions with pff grades,} \emph{Pro Football Focus Research and
  Development Journal}, 1.

\bibitem[{Drinen(2013)}]{Drinen}
Drinen, D. (2013): \enquote{Approximate value: Methodolgy,}
  \urlprefix\url{https://www.sports-reference.com/blog/approximate-value-methodology/}.

\bibitem[{Eager et~al.(2017)Eager, Chahrouri, and Palazzolo}]{Eager17}
Eager, E.~A., G.~Chahrouri, and S.~Palazzolo (2017): \enquote{Using pff grades
  to cluster quarterback performance,} \emph{Pro Football Focus Research and
  Development Journal}, 1.

\bibitem[{Elmore and DeWitt(2017)}]{Elmore17}
Elmore, R. and P.~DeWitt (2017): \emph{ballr: Access to Current and Historical
  Basketball Data}, \urlprefix\url{https://CRAN.R-project.org/package=ballr}, r
  package version 0.1.1.

\bibitem[{ESPN(2017)}]{ESPN_fantasy}
ESPN (2017): \enquote{Scoring formats,}
  \urlprefix\url{http://games.espn.com/ffl/resources/help/content?name=scoring-formats}.

\bibitem[{Franks et~al.(2017)Franks, D'Amour, Cervone, and Bornn}]{Franks17}
Franks, A.~M., A.~D'Amour, D.~Cervone, and L.~Bornn (2017):
  \enquote{Meta-analytics: tools for understanding the statistical properties
  of sports metrics,} \emph{Journal of Quantitative Analysis in Sports}, 4.

\bibitem[{Gelman and Hill(2007)}]{Gelman07}
Gelman, A. and J.~Hill (2007): \emph{Data Analysis Using Regression and
  Multilevel/Hierarchical Models}, Cambridge, United Kingdom: Cambridge
  University Press.

\bibitem[{Goldner(2012)}]{Goldner12}
Goldner, K. (2012): \enquote{A markov model of football: Using stochastic
  processes to model a football drive,} \emph{Journal of Quantitative Analysis
  in Sports}, 8.

\bibitem[{Goldner(2017)}]{Goldner17}
Goldner, K. (2017): \enquote{Situational success: Evaluating decision-making in
  football,} in J.~Albert, M.~E. Glickman, T.~B. Swartz, and R.~H. Koning,
  eds., \emph{Handbook of Statistical Methods and Analyses in Sports}, Boca
  Raton, Florida: CRC Press, 183--198.

\bibitem[{Gramacy et~al.(2012)Gramacy, Taddy, and Jensen}]{Gramacy12}
Gramacy, R.~B., M.~A. Taddy, and S.~T. Jensen (2012): \enquote{Estimating
  player contribution in hockey with regularized logistic regression,}
  \emph{Journal of Quantitative Analysis in Sports}, 9.

\bibitem[{Grimshaw and Burwell(2014)}]{Grimshaw14}
Grimshaw, S.~D. and S.~J. Burwell (2014): \enquote{Choosing the most popular
  nfl games in a local tv market,} \emph{Journal of Quantitative Analysis in
  Sports}, 10.

\bibitem[{Horowitz et~al.(2017)Horowitz, Yurko, and Ventura}]{Horowitz17}
Horowitz, M., R.~Yurko, and S.~L. Ventura (2017): \emph{nflscrapR: Compiling
  the NFL play-by-play API for easy use in R},
  \urlprefix\url{https://github.com/maksimhorowitz/nflscrapR}, r package
  version 1.4.0.

\bibitem[{James(2017)}]{James17}
James, B. (2017): \enquote{Judge and altuve,}
  \urlprefix\url{https://www.billjamesonline.com/judge\_and\_altuve/}.

\bibitem[{Jensen and Turner(2014)}]{Jensen14}
Jensen, J.~A. and B.~A. Turner (2014): \enquote{What if statisticians ran
  college football? a re-conceptualization of the football bowl subdivision,}
  \emph{Journal of Quantitative Analysis in Sports}, 10.

\bibitem[{Jensen et~al.(2009)Jensen, Shirley, and Wyner}]{Jensen09}
Jensen, S., K.~E. Shirley, and A.~Wyner (2009): \enquote{Bayesball: A bayesian
  hierarchical model for evaluating fielding in major league baseball,}
  \emph{The Annals of Applied Statistics}, 3.

\bibitem[{Judge et~al.(2015{\natexlab{a}})Judge, Brooks, and
  Pavlidis}]{Brooks15}
Judge, J., D.~Brooks, and H.~Pavlidis (2015{\natexlab{a}}): \enquote{Moving
  beyond wowy: A mixed approach to measuring catcher framing,}
  \urlprefix\url{https://www.baseballprospectus.com/news/article/25514/moving-beyond-wowy-a-mixed-approach-to-measuring-catcher-framing/}.

\bibitem[{Judge et~al.(2015{\natexlab{b}})Judge, Turkenkopf, and
  Pavlidis}]{Turkenkopf15}
Judge, J., D.~Turkenkopf, and H.~Pavlidis (2015{\natexlab{b}}):
  \enquote{Prospectus feature: Introducing deserved run average (dra) and all
  its friends,}
  \urlprefix\url{https://www.baseballprospectus.com/news/article/26195/prospectus-feature-introducing-deserved-run-average-draand-all-its-friends/}.

\bibitem[{Katz and Burke(2017)}]{ESPN_total_QBR}
Katz, S. and B.~Burke (2017): \enquote{How is total qbr calculated? we explain
  our quarterback rating,}
  \urlprefix\url{http://www.espn.com/blog/statsinfo/post/\_/id/123701/how-is-total-qbr-calculated-we-explain-our-quarterback-rating}.

\bibitem[{Kubatko et~al.(2007)Kubatko, Oliver, Pelton, and
  Rosenbaum}]{Kubatko07}
Kubatko, J., D.~Oliver, K.~Pelton, and D.~T. Rosenbaum (2007): \enquote{A
  starting point for analyzing basketball statistics,} \emph{Journal of
  Quantitative Analysis in Sports}, 3.

\bibitem[{Lahman(1996 -- 2017)}]{Lahman}
Lahman, S. (1996 -- 2017): \emph{Lahman's Baseball Database},
  \urlprefix\url{http://www.seanlahman.com/baseball-archive/statistics/}.

\bibitem[{Lillibridge(2013)}]{Lillibridge13}
Lillibridge, M. (2013): \enquote{The anatomy of a 53-man roster in the nfl,}
  \urlprefix\url{http://bleacherreport.com/articles/1640782-the-anatomy-of-a-53-man-roster-in-the-nfl}.

\bibitem[{Lock and Nettleton(2014)}]{Lock14}
Lock, D. and D.~Nettleton (2014): \enquote{Using random forests to estimate win
  probability before each play of an nfl game,} \emph{Journal of Quantitative
  Analysis in Sports}, 10.

\bibitem[{Lopez(2017)}]{Lopez17}
Lopez, M. (2017): \enquote{All win probability models are wrong some are
  useful,}
  \urlprefix\url{https://statsbylopez.com/2017/03/08/all-win-probability-models-are-wrong-some-are-useful/}.

\bibitem[{Macdonald(2011)}]{Macdonald11}
Macdonald, B. (2011): \enquote{A regression-based adjusted plus-minus statistic
  for nhl players,} \emph{Journal of Quantitative Analysis in Sports}, 7.

\bibitem[{Martin et~al.(2017)Martin, Timmons, and Powell}]{Martin17}
Martin, R., L.~Timmons, and M.~Powell (2017): \enquote{A markov chain analysis
  of nfl overtime rules,} \emph{Journal of Sports Analytics}, Pre-print.

\bibitem[{Meers(2011)}]{Meers11}
Meers, K. (2011): \enquote{How to value nfl draft picks,}
  \urlprefix\url{https://harvardsportsanalysis.wordpress.com/2011/11/30/how-to-value-nfl-draft-picks/}.

\bibitem[{Morris(2015)}]{Morris15}
Morris, B. (2015): \enquote{Kickers are forever,}
  \urlprefix\url{https://fivethirtyeight.com/features/kickers-are-forever/}.

\bibitem[{Morris(2017)}]{Morris17}
Morris, B. (2017): \enquote{Running backs are finally getting paid what
  they’re worth,}
  \urlprefix\url{https://fivethirtyeight.com/features/running-backs-are-finally-getting-paid-what-theyre-worth/}.

\bibitem[{Mulholland and Jensen(2014)}]{Mulholland14}
Mulholland, J. and S.~T. Jensen (2014): \enquote{Predicting the draft and
  career success of tight ends in the national football league,} \emph{Journal
  of Quantitative Analysis in Sports}, 10.

\bibitem[{Oliver(2011)}]{Oliver11}
Oliver, D. (2011): \enquote{Guide to the total quarterback rating,}
  \urlprefix\url{http://www.espn.com/nfl/story/\_/id/6833215/explaining-statistics-total-quarterback-rating}.

\bibitem[{Paine(2015)}]{Paine15b}
Paine, N. (2015): \enquote{Bryce harper should have made \$73 million more,}
  \urlprefix\url{https://fivethirtyeight.com/features/bryce-harper-nl-mvp-mlb/}.

\bibitem[{Pasteur and Cunningham-Rhoads(2014)}]{Pasteur14}
Pasteur, D. and K.~Cunningham-Rhoads (2014): \enquote{An expectation-based
  metric for nfl field goal kickers,} \emph{Journal of Quantitative Analysis in
  Sports}, 10.

\bibitem[{Pattani(2012)}]{ESPN_EP}
Pattani, A. (2012): \enquote{Expected points and epa explained,}
  \urlprefix\url{http://www.espn.com/nfl/story/\_/id/8379024/nfl-explaining-expected-points-metric}.

\bibitem[{Phillips(2018)}]{Phillips}
Phillips, C. (2018): \emph{Scoring and Scouting: How We Know What We Know about
  Baseball}, Princeton, New Jersey: Forthcoming, Princeton University Press.

\bibitem[{Piette and Jensen(2012)}]{Piette12}
Piette, J. and S.~Jensen (2012): \enquote{Estimating fielding ability in
  baseball players over time,} \emph{Journal of Quantitative Analysis in
  Sports}, 8.

\bibitem[{Pro-Football-Reference(2018)}]{PFR}
Pro-Football-Reference (2018): \enquote{Football glossary and football
  statistics glossary,}
  \urlprefix\url{https://www.pro-football-reference.com/about/glossary.htm}.

\bibitem[{Quealy et~al.(2017)Quealy, Causey, and Burke}]{Quealy}
Quealy, K., T.~Causey, and B.~Burke (2017): \enquote{4th down bot: Live
  analysis of every n.f.l. 4th down,}
  \urlprefix\url{http://nyt4thdownbot.com/}.

\bibitem[{{R Core Team}(2017)}]{R17}
{R Core Team} (2017): \emph{R: A Language and Environment for Statistical
  Computing}, R Foundation for Statistical Computing, Vienna, Austria,
  \urlprefix\url{https://www.R-project.org/}.

\bibitem[{Ratcliffe(2013)}]{PFF_fantasy}
Ratcliffe, J. (2013): \enquote{The pff idp scoring system revisited,}
  \urlprefix\url{https://www.profootballfocus.com/news/the-pff-idp-scoring-system-revisited}.

\bibitem[{Romer(2006)}]{Romer06}
Romer, D. (2006): \enquote{Do firms maximize? evidence from professional
  football,} \emph{Journal of Political Economy}, 114.

\bibitem[{Rosenbaum(2004)}]{Rosenbaum04}
Rosenbaum, D.~T. (2004): \enquote{Measuring how nba players help their teams
  win,} \urlprefix\url{http://www.82games.com/comm30.htm}.

\bibitem[{Schatz(2003)}]{Schatz03}
Schatz, A. (2003): \enquote{Methods to our madness,}
  \urlprefix\url{http://www.footballoutsiders.com/info/methods}.

\bibitem[{Schoenfield(2012)}]{Schoenfield12}
Schoenfield, D. (2012): \enquote{What we talk about when we talk about war,}
  \urlprefix\url{http://espn.go.com/
  blog/sweetspot/post/_/id/27050/what-we-talk-about-when-we-talk-about-war}.

\bibitem[{Sievert(2015)}]{Sievert15}
Sievert, C. (2015): \emph{pitchRx: Tools for Harnessing 'MLBAM' 'Gameday' Data
  and Visualizing 'pitchfx'},
  \urlprefix\url{https://CRAN.R-project.org/package=pitchRx}, r package version
  1.8.2.

\bibitem[{Skiver(2017)}]{CBS}
Skiver, K. (2017): \enquote{Zebra technologies to bring live football tracking
  on nfl gamedays with rfid chips,}
  \urlprefix\url{https://www.cbssports.com/nfl/news/zebra-technologies-to-bring-live-football-tracking-on-nfl-gamedays-with-rfid-chips/}.

\bibitem[{Smith et~al.(1973)Smith, Siwoff, and Weiss}]{Smith73}
Smith, D., S.~Siwoff, and D.~Weiss (1973): \enquote{Nfl's passer rating,}
  \urlprefix\url{http://www.profootballhof.com/news/nfl-s-passer-rating}.

\bibitem[{Snyder and Lopez(2015)}]{Snyder15}
Snyder, K. and M.~Lopez (2015): \enquote{Consistency, accuracy, and fairness: a
  study of discretionary penalties in the nfl,} \emph{Journal of Quantitative
  Analysis in Sports}, 11.

\bibitem[{Sportradar(2017)}]{sportradar_agreement}
Sportradar (2017): \emph{NFL Data Addendum},
  \urlprefix\url{https://developer.sportradar.com/page/NFL_Addendum}, last
  updated 2017-08-11.

\bibitem[{Sportradar(2018)}]{sportradar_data}
Sportradar (2018): \emph{NFL Official API},
  \urlprefix\url{https://developer.sportradar.com/files/indexFootball.html#player-participation},
  version 2.

\bibitem[{Tango(2017)}]{Tango17}
Tango, T. (2017): \enquote{War podcast,}
  \urlprefix\url{http://tangotiger.com/index.php/site/comments/war-podcast}.

\bibitem[{Tango et~al.(2007)Tango, Lichtman, and Dolphin}]{Tango07}
Tango, T., M.~Lichtman, and A.~Dolphin (2007): \emph{The Book: Playing the
  Percentages in Baseball}, Washington, D.C: Potomac Book, Inc.

\bibitem[{Thomas and Ventura(2017)}]{Thomas17}
Thomas, A. and S.~L. Ventura (2017): \emph{nhlscrapr: Compiling the NHL Real
  Time Scoring System Database for easy use in R},
  \urlprefix\url{https://CRAN.R-project.org/package=nhlscrapr}, r package
  version 1.8.1.

\bibitem[{Thomas and Ventura(2015)}]{Thomas15}
Thomas, A.~C. and S.~L. Ventura (2015): \enquote{The road to war,}
  \urlprefix\url{http://blog.war-on-ice.com/index.html\%3Fp=429.html}.

\bibitem[{Thomas et~al.(2013)Thomas, Ventura, Jensen, and Ma}]{Thomas13}
Thomas, A.~C., S.~L. Ventura, S.~T. Jensen, and S.~Ma (2013):
  \enquote{Competing process hazard function models for player ratings in ice
  hockey,} \emph{The Annals of Applied Statistics}, 7.

\bibitem[{Wakefield and Rivers(2012)}]{Wakefield12}
Wakefield, K. and A.~Rivers (2012): \enquote{The effect of fan passion and
  official league sponsorship on brand metrics: A longitudinal study of
  official nfl sponsors and roo,} \emph{MIT Sloan Sports Analytics Conference}.

\bibitem[{Yam and Lopez(2018)}]{Yam18}
Yam, D.~R. and M.~J. Lopez (2018): \enquote{Quantifying the causal effects of
  conservative fourth down decision making in the national football league.}
  \urlprefix\url{https://statsbylopez.files.wordpress.com/2018/01/quantifying-causal-effects.pdf},
  under Review.

\bibitem[{Yurko et~al.(2017)Yurko, Ventura, and Horowitz}]{Yurko17}
Yurko, R., S.~Ventura, and M.~Horowitz (2017): \enquote{Nfl player evaluation
  using expected points added with nflscrapr,} Presented at the Great Lakes
  Sports Analytics Conference.

\bibitem[{Zhou and Ventura(2017)}]{Zhou17}
Zhou, E. and S.~Ventura (2017): \enquote{Wins and point differential in the
  nfl,}
  \urlprefix\url{https://www.cmusportsanalytics.com/wins-point-differential-nfl/}.

\end{thebibliography}

\end{document}